\documentclass[aps,prx,twocolumn,amsmath,nofootinbib,longbibliography,amssymb,superscriptaddress,10pt]{revtex4-2}

\usepackage{tikz}
\usepackage{tikz-feynman}
\usepackage{amsmath}
\usepackage{relsize}
\tikzfeynmanset{compat=1.1.0}
\usepackage[english]{babel}
\usepackage{graphicx}
\usepackage{braket}
\usepackage[normalem]{ulem}
\usepackage{soul}
\usepackage{float} 
\usepackage[svgnames]{xcolor} 
\usepackage[colorlinks=true, linkcolor=Red , citecolor= NavyBlue,urlcolor = NavyBlue, breaklinks=true]{hyperref}
\usepackage{verbatim}
\usepackage{esint}
\usepackage{marginnote}
\usepackage{graphicx,mathtools,bm,bbm}
\usepackage{braket}

\def \k {\mathbf{k}}
\def \q {\mathbf{q}}

\def \R {\mathbf{R}}

\def \a {\mathbf{a}}
\def \b {\mathbf{b}}
\def \sgn {\operatorname{sgn}}

\begin{document}
\title{Controlled Loop Expansion for the Topological Heavy Fermion Model}
\author{Yaar Vituri}
\affiliation{Department of Condensed Matter Physics, Weizmann Institute of Science, Rehovot 76100, Israel}

\author{Erez Berg}
\affiliation{Department of Condensed Matter Physics, Weizmann Institute of Science, Rehovot 76100, Israel}
\affiliation{Materials Department, University of California Santa Barbara, Santa Barbara 93106 USA}
\affiliation{Department of Electrical and Computer Engineering, University of California, Santa Barbara, CA 93106,
USA}

\date{\today}
\begin{abstract}
    We develop a controlled theoretical framework for the topological heavy fermion model relevant to magic-angle twisted bilayer graphene, where low density conduction electrons hybridize with a lattice of strongly interacting $f$-sites. By tracing out the localized electrons, we derive an effective action for the conduction electrons with long-range in time effective interactions, built from correlators of the single $f$-site problem. We identify a small hybridization-phase-space parameter resulting in a controlled loop expansion, enabling the derivation of nonperturbative results in either the interaction or the hybridization strength. To tree-level, the results are equivalent to the Hubbard I approximation. At higher loop order, we derive two key results applicable to temperatures above the flavor ordering temperature and below the on-site charging energy: 1) the quasi-particle lifetime, 2) the flavor susceptibility of the system. Remarkably, despite being strongly interacting, we find the susceptibility to accurately obey a Curie-Weiss law parametrically close to the Curie temperature. 

\end{abstract}
\maketitle

\section{Introduction}
In recent years, twisted Van der Waals (VdW) heterostructures have emerged as versatile platforms for exploring strongly correlated electronic phenomena. These systems host a remarkable range of many-body phenomena, including correlated insulators~\cite{cao2018correlated, lu2019superconductors, codecido2019correlated, saito2020independent, pierce2021unconventional, cao2020tunable,  kim2023imaging}, superconductivity~\cite{cao2018unconventional, yankowitz2019tuning, lu2019superconductors, codecido2019correlated, saito2020independent, cao2021nematicity, tian2023evidence, stepanov2020untying, arora2020superconductivity, liu2020tunable, hao2021electric, park2022robust, cao2021pauli, banerjee2025superfluid, kim2022evidence, kim2026resolving}, Pomeranchuk effect~\cite{zondiner2020cascade, saito2021isospin, rozen2021entropic, zhang2025heavy, choi2019electronic}, linear-in-$T$ resistivity~\cite{polshyn2019large, cao2020strange, jaoui2022quantum}, Chern insulators~\cite{grover2022chern, wu2021chern, xie2021fractional, pierce2021unconventional}, and both integer and fractional quantum anomalous Hall states~\cite{sharpe2019emergent, serlin2020intrinsic, stepanov2021competing}. Among these, magic angle twisted bilayer graphene (MATBG) stands out. In addition to realizing the majority of the aforementioned phenomena, MATBG has been studied extensively, both theoretically~\cite{bistritzer2011moire, wu2018theory, tarnopolsky2019origin, po2019faithful, hofmann2022fermionic, khalaf2021charged, xie2020nature, bultinck2020ground, kwan2021kekule, wagner2022global, shavit2021theory, kwan2024electron, antebi2022plane, mendez2024low, lewandowski2021pairing, ledwith2020fractional, khalaf2019magic, khalaf2020soft, parker2021field, liu2021nematic, ledwith2022family, ledwith2024nonlocal,kumar2021lattice,song2019all,kang2018symmetry,bernevig2021twisted_III, bernevig2021twisted_V, lian2021twisted, herzog2025topological, datta2023heavy} and experimentally~\cite{xiao2025interactingenergybandsmagic, cao2018correlated,cao2018unconventional,lu2019superconductors,codecido2019correlated,saito2020independent,cao2020strange,pierce2021unconventional,yankowitz2019tuning,cao2021nematicity,tian2023evidence,stepanov2020untying,stepanov2021competing,zondiner2020cascade,arora2020superconductivity,rozen2021entropic,zhang2025heavy,choi2019electronic,polshyn2019large,jaoui2022quantum,grover2022chern,wu2021chern,xie2021fractional,sharpe2019emergent,serlin2020intrinsic,saito2021isospin}, establishing a detailed picture of its phase diagram. 

Despite extensive efforts, analytical understanding of MATBG remains limited. Early theoretical progress was largely restricted to zero-temperature and integer fillings, where the ground state is well approximated by a single Slater determinant~\cite{bultinck2020ground,kwan2021kekule,wagner2022global,shavit2021theory,kwan2024electron,xie2020nature, lian2021twisted}.
Developing a controlled description of the system at finite temperatures and away from integer filling
is essential for understanding the metallic properties and the origins of superconductivity in MATBG.

More recently, significant progress has been made toward describing these finite-temperature and finite-doping regimes. A key conceptual step was provided by Song and Bernevig~\cite{song2022magic}, who showed that the low-energy bands of the Bistritzer–MacDonald model~\cite{bistritzer2011moire} -- the widely accepted single-particle description of MATBG -- can be recast in terms of localized $f$-electrons hybridized with itinerant, Dirac-like $c$-electrons. 

This basis change, referred to as the topological heavy fermion (THF) basis, situates MATBG within the broader context of heavy-fermion physics.
It offers a natural framework for understanding the interplay between topology and localized-electron behavior, and provides an interpretation of the compressibility features
~\cite{zondiner2020cascade,rozen2021entropic,saito2021isospin}
as signatures of charge redistribution between $f$ and $c$ electrons. This interpretation was corroborated numerically using dynamical mean-field theory (DMFT)~\cite{georges1996dynamical, rai2024dynamical, crippa2025dynamical}.

The THF basis on its own does not yield an immediate path to analytically controlled results. This limitation mirrors a longstanding challenge in finding an accurate and controlled description of heavy-fermion physics. The lack of a small parameter and the intricate interplay between local and itinerant states generally preclude controlled analytic results~\cite{coleman2006heavy,hewson1993kondo,lohneysen2007fermi}. 

A promising step toward an analytical approach was provided by Ledwith \textit{et al.}~\cite{ledwith2024nonlocal}, who identified a small parameter in the Hamiltonian description of MATBG.
This parameter, denoted by $s^2$, characterizes the small fraction of the Brillouin zone having nontrivial quantum geometry in the low-energy bands. For MATBG this parameter is estimated to be about $5$-$10\%$ of the Brillouin zone area. In the THF basis this parameter can be understood as the fraction of momentum space where the Bloch wave functions of the low-energy bands retain appreciable $c$-electron character~\cite{hu2025projected}. Ref.~\cite{ledwith2024nonlocal} derived the electronic spectral function in the limit of $s^2\rightarrow 0$, and estimated the entropy per moir\'{e} site at low but nonzero temperatures.

In this work, we introduce a systematic diagrammatic loop expansion for the THF model, inspired by the small parameter of Ref.~\cite{ledwith2024nonlocal}. Organizing diagrams by loop order directly identifies the small parameters governing the expansion, establishing control without invoking the chiral or projected limits.  Our framework therefore applies without assuming any hierarchy between the gap separating the narrow and remote bands and the interaction scale.

We identify two key small parameters, $\mathcal{I}/U\ll 1$ and $\Theta_c/T\ll 1$, 
with $\mathcal{I}$ being the hybridization function ($\operatorname{Im}[\mathcal{I}]$ is the $c$ electron density of states, weighted by the $c$-$f$ hybridization), $U$ is the on-site (Hubbard) interaction on the $f$-sites, $\Theta_c$ is the Curie temperature and $T$ is the temperature. In the chiral and projected limit these small parameters are related to $s^2$ through $\mathcal{I}(\omega)\sim s^2N_f\omega$ and $\Theta_c\sim s^2U$ (up to logarithmic corrections). 
For the diagrams we consider we find that each loop in a given diagram contains a factor of $\frac{\mathcal{I}}{U}$ or  $\frac{\Theta_c}{T}$. In the flat-chiral limit we argue that for any diagram, each loop contains a factor of $s^2$ thus making the loop expansion a controlled approximation.

Crucially, in the chiral and projected limit of MATBG, the small parameter required to control the calculations is 
\begin{equation}
    s^2N_f\ll1~.
\end{equation}
 It differs from the small parameter argued for in Refs.~\cite{ledwith2024nonlocal, hu2025projected} by a factor of $N_f$. For realistic parameters for MATBG we find the control parameter to be only moderately small, $s^2N_f\approx 0.5$. Nevertheless, we believe that studying the system in the controlled limit will capture key aspects of MATBG.

To zeroth order in the small parameters, our results reproduce those of Ref.~\cite{ledwith2024nonlocal} in the chiral and projected limit. Away from these limits, the zeroth order corresponds to the Hubbard-I approximation~\cite{hubbard1963electron,hu2025projected}. 

By going to finite loop order we are able to derive the quasiparticle lifetime and the temperature dependent flavor susceptibility of the system. 
The Mott bands are found to have a significant broadening, of the order of $s^2N_fU$.
We find the flavor susceptibility to have a Curie-Weiss form, diverging at the Curie temperature $\Theta_c \sim U s^2$.

The rest of this paper is organized as follows. In Sec.~\ref{section:summary} we provide a brief summary of the main physical results. In Sec.~\ref{section:model} we define the model and describe its application to MATBG. In Sec.~\ref{section:expansion} we derive the effective action for the $c$ electrons and establish the controlled loop expansion. In Sec.~\ref{section:QP}, we derive the quasiparticle self-energy at one-loop order and beyond, including an infinite resummation near the Mott bands. 
Sec.~\ref{section:susceptibility} derives the Curie-Weiss form of the flavor susceptibility. 
We conclude with a discussion and outlook in Sec.~\ref{section:discussion}.

\begin{figure*}
    \centering
    \includegraphics[width=1.0\linewidth]{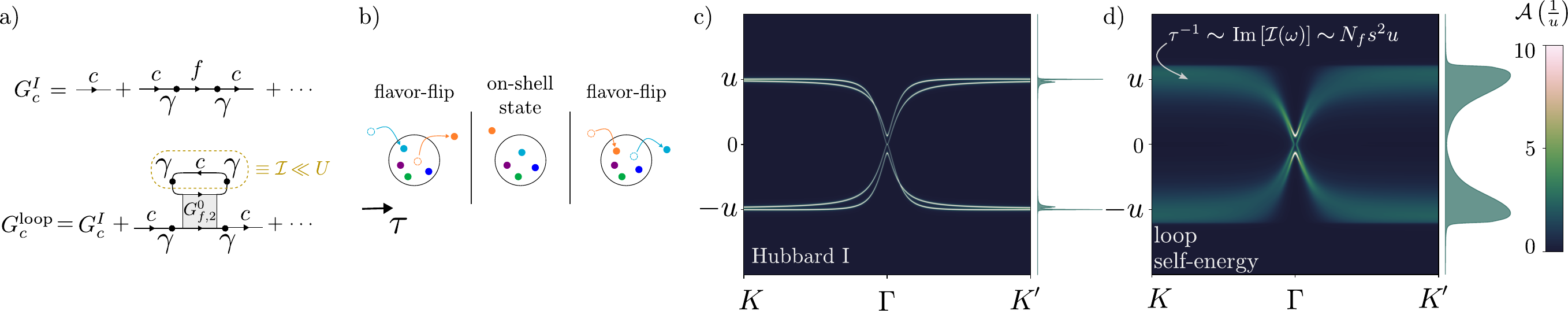}
    \caption{
    a) A diagrammatic representation of the $c$ electron propagator within Hubbard I (tree-level) approximation ($G_c^I$) and to one-loop order ($G_c^\text{loop}$), expressed diagrammatically in terms of the un-hybridized ($\hat{\gamma}=0$) theory. 
    $G_{f,2}^0$ represents the connected two-body Green's function of the $f$ electrons. The one loop self-energy encodes information about correlation in the $f$-site beyond its single-particle propagator. $\mathcal{I}$ is the hybridization integral defined in Eq.~\eqref{eq:hybrid_int}, with $\mathcal{I}/U$ serving as the control parameter for our loop expansion. 
    b) The flavor-flip process, corresponding to $G_{f,2}^0$ at specific time orderings. The on-shell (zero energy) state at intermediate times results in an effective elastic scattering for low energy $c$ electrons. 
    c,d) The spectral function $\mathcal{A}(\k,\omega)$ at charge neutrality ($\nu=0$) within Hubbard I approximation (c); and within the loop resummation derived in this work (d). To the right of each panel is the local (momentum integrated) spectral function. The Hubbard I spectral function is characterized by infinitely sharp excitations as an artifact of including only tree-level diagrams in the effective theory. Away from the $\Gamma$-point the excitations are dominantly $f$ like and their scattering rate $\tau^{-1}$ is given by the imaginary part of the hybridization integral. 
    }
    \label{fig:header}
\end{figure*}

\section{Summary of Results}
\label{section:summary}

Our starting point is a periodic Anderson model of itinerant ($c$) and localized ($f$) electrons with local interactions on the $f$ sites. We trace out the $f$ electrons with respect to the single-site action, resulting in an effective theory for the $c$-electrons with extended-in-time vertices [see Eq.~\eqref{eq:eff_action}].

By studying the two-body vertex and the one-loop self-energy derived from it we identify flavor-flip processes as the dominant ones,
resulting in a finite quasiparticle lifetime at one-loop order (Fig.~\ref{fig:header}.a and .b).
Other processes generate short-range in time interactions, affecting the quasiparticle lifetime only at higher order in the loop expansion.

We define the hybridization function $\mathcal{I}(\omega)$ in Eq.~\eqref{eq:hybrid_int} as the local $c$-electron Green's function weighted by the $c$-$f$ hybridization, analogous to the hybridization function used in dynamical mean-field theory (DMFT). We find that in order to control the magnitude of the one-loop contribution relative to the tree-level (zero loops) self-energy, we require the hybridization function to be small compared to the interaction scale $\frac{|\mathcal{I}(\omega)|}{U}\ll 1$.

In the flat-chiral and projected limit of the THF model we show that $\mathcal{I}(\omega)\sim s^2N_f\omega$ up to logarithmic corrections, with $s$ being the small parameter identified in Ref.~\cite{ledwith2024nonlocal}. In this limit, any diagram with $n_l$ loops in our effective theory is suppressed by at most $( s^2N_f)^{n_l}$.  Therefore, we find a necessary condition for the control of the expansion in the flat-chiral limit to be $s^2N_f\ll1$. This is a stricter condition than the condition of $s^2\ll1$ assumed to be sufficient in previous works. For unstrained MATBG $N_f=8$, corresponding to spin, valley, and orbital degrees of freedom. Realistic parameters give $s^2\approx\frac{1}{16}$, making the control parameter of the expansion only marginally small.\footnote{In the explicit calculations presented here we choose parameters such that $s^2=1/64$, in order to be well inside the controlled regime.}

Within our expansion, to tree-level (zeroth order in number of loops), the one-body propagator is identical to the one in the Hubbard I approximation~\cite{hubbard1963electron} (Fig.~\ref{fig:header}.c), where the spectral function consists of infinitely sharp $\delta$-functions. In the chiral and projected limit, the tree-level propagator reproduces the results of Ledwith et al.~\cite{ledwith2024nonlocal}. At one-loop order (Sec.~\ref{section:one loop}) we find that the quasiparticles gain a finite lifetime. 

For quasiparticles within the Mott bands the one-loop diagram diverges as $\frac{\mathcal{I}}{\omega-\omega_{\text{Mott}}}$. To derive a regular expression, we must sum all higher-loop diagrams which contain higher powers of the same divergent term. In Sec.~\ref{section:loop-resummation} we show that the appropriate resummation can be done analytically in the limit of many $f$ electron flavors (large $N_f$). We find a significant broadening of the Mott bands, of the order of $\operatorname{Im}\mathcal{I}(\omega_{\text{Mott}}) \sim s^2N_fU$ (Fig.~\ref{fig:header}.d).

In Sec.~\ref{section:excluded} we go beyond strictly spatially local diagrams. We calculate the lowest loop-order non-local diagram contributing to the self-energy. Similarly to the one-loop diagram, we find this diagram to be of order $\frac{|\mathcal{I}(\omega)|}{U}$. Relative to the one-loop diagram, the leading non-local diagram has an additional factor of $\frac{\Theta_c}{T}$, with $\Theta_c$ defined in Eq.~\eqref{eq:Theta_c}. In the flat-chiral and projected limit we find $\Theta_c\sim Us^2$. At elevated temperatures, $T\gg\Theta_c$, the factor $\frac{\Theta_c}{T}$ allows us to neglect the effect of non-local diagrams, whereas at low temperatures of order $\Theta_c$ it results in the breakdown of the approximation.

Within the limits of validity of our theory, the electronic self-energy is found to be essentially spatially local. This occurs because all leading-order diagrams in the resummation are local, with non-local diagrams suppressed by powers of $\tfrac{\Theta_c}{T}$. 
Thus, at intermediate temperatures ($\Theta_c\ll T \ll U$), we establish a direct connection between our approach and DMFT~\cite{georges1996dynamical, gull2011continuous}, where the self-energy is assumed to be local.

Finally, to better understand how our approach breaks down at low temperature, we study the flavor susceptibility of the model in Sec.~\ref{section:susceptibility}. 
We find that the susceptibility obeys a Curie-Weiss scaling, 
\begin{equation}
    \chi\sim\frac{1}{T-\Theta_c}~,
\end{equation} 
down to parametrically small reduced temperatures above $\Theta_c$. 
The Curie-Weiss form of the susceptibility remains accurate parametrically close to the Curie temperature, down to $T-\Theta_c\gtrsim s^2\Theta_c\sim s^4U$.
The Curie temperature is identical to the energy scale appearing in the low temperature divergence of the non-local self-energy diagram, indicating that the appearance of finite-range flavor correlations is the cause for the breakdown of the locality of the self-energy at low temperatures. 

Finally, we study the nature of the leading flavor ordering instability. In the flat-chiral limit we find the flavor susceptibility to be $\mathrm{SU}(N_f)$ symmetric, where dispersion and deviation from the chiral limit lift different degeneracies. The resulting hierarchy of nearly degenerate orders is similar to that of Ref.~\cite{bultinck2020ground}, with an enhanced symmetry at the flat-chiral limit.

\section{Model}
\label{section:model}
As a starting point, we consider a general multi-orbital periodic Anderson model. The action is given by 
\begin{equation}
    S=S_c+S_f+S_{fc}~,
    \label{eq:action}
\end{equation}
with
\begin{align}
    S_c &= \int_\tau \sum_{\k,\lambda} \bar{c}_{\k,\lambda,a} \left[(\partial_\tau -\mu)\delta_{aa'}+ h^{(c,\lambda)}_{\k,aa'} \right]c_{\k,\lambda,a'}~,  \\
    S_f &= \int_\tau \sum_{\R} \left[
    \bar{f}_{\R,\lambda,b}  \left(\partial_\tau -\mu\right)f_{\R,\lambda,b} +\frac{U}{2} \delta n_\R^2 \right]~, \\
    S_{fc} &=\int_\tau \sum_{\k,\lambda} (\hat{\gamma}_{\k})_{ba}\left(\sum_{\R}\frac{e^{i\k\cdot \R} }{\sqrt{N_{s}}}\bar{f}_{\R,\lambda,b}\right)c_{\k,\lambda,a} + \operatorname{h.c.}~,
\end{align}
where $\int_\tau \equiv \int_0^\beta d\tau$, $\delta n_\R\equiv\sum_{\lambda,b}\left( \bar{f}_{\R,\lambda,b}f_{\R,\lambda,b}-\frac{1}{2}\right)$ is the $f$-site charge deviation from neutrality.
We use $a,a'$ to denote the $c$ orbitals and $b,b'$ to denote the $f$ orbitals. $\lambda$ denotes the flavor (spin and valley) index. In the topological heavy fermion model~\cite{song2022magic} the numbers of orbitals and flavors are $N_a=4, N_b=2, N_\lambda=4$ with $\lambda=(s,\tau)$ for spin and valley, respectively. $\k$ denotes summation over real (unrestricted) momenta, $\R$ denotes moir\'{e} lattice positions, and $N_s$ is the number of lattice sites. The matrix $(\hat{\gamma}_{\k})_{ba}$ encodes the hybridization between the $f$ and $c$ electrons. 

For simplicity, in the following we restrict $\k$ to be in the first Brillouin zone.
This assumption is non-essential, and the expansion could be carried out without it. It simplifies the notation by allowing us to define $c_{\R,\lambda,a}$ and $\hat{\gamma}_{ba}(\R-\R')$ such that 
\begin{equation}
    c_{\k,\lambda,a} = \sum_\R \frac{e^{-i\k\cdot \R} }{\sqrt{N_{s}}}c_{\R,\lambda,a}
\end{equation}
for every $\k$ in the first Brillouin zone, and 
\begin{equation}
    \hat{\gamma}_{ba}(\R-\R') = \sum_\k \frac{e^{i\k\cdot(\R-\R')}}{N_s}(\hat{\gamma}_{\k})_{ba}~.
\end{equation}
We can now express $S_{fc}$ in real space as a non-local hybridization term:
\begin{equation}
    S_{fc} =\int_0^\beta d\tau \sum_{\R,\R',\lambda} \hat{\gamma}_{ba}(\R-\R')\bar{f}_{\R,\lambda,b}c_{\R',\lambda,a} + \operatorname{h.c.}~.
\end{equation}

\subsection{Application to MATBG}
We now focus on the application of the model to MATBG. 
We use the Song-Bernevig Topological Heavy Fermion (THF) model~\cite{song2022magic} that faithfully represents MATBG. 
Within this model we have
\begin{equation}
    \hat{h}_\k^{(c,\lambda)} = 
    \left( \begin{array}{cc}
         0_{2\times2} &  v_\star\sigma_x(\k^{(\lambda)}\cdot\vec{\sigma})\\
        v_\star(\k^{(\lambda)}\cdot\vec{\sigma})\sigma_x & M\sigma_x
    \end{array} \right),
    \label{eq:H_c}
\end{equation}
and
\begin{equation}
    \hat{\gamma}_\k = \left(\gamma\sigma_0+v'_\star\k^{(\lambda)}\cdot\vec{\sigma},~0_{2\times2}
    \right),
    \label{eq:gamma_def}
\end{equation}
where $\k^{(\lambda)}=(\tau_zk_x,k_y)$, with $\tau_z$ being the valley index, and $\sigma_{x,y,z}$ are Pauli matrices acting in orbital space. 
For brevity, we introduce another set of Pauli matrices $(\zeta_0,\zeta_x,\zeta_y,\zeta_z)$ which act on the remaining orbital degree of freedom of the $c$ electrons, and commute with $\sigma_{x,y,z}$, such that
\begin{align}
    \hat{h}_\k^{(c,\lambda)} &= \tau_z\sigma_0k_x\zeta_x-\tau_0\sigma_zk_y\zeta_y +M\sigma_x\left(\frac{\zeta_0-\zeta_z}{2}\right).
    \label{eq:reduced_Hc}
\end{align}
Further details of the model are provided in App.~\ref{App:THF_model}. 
The THF model includes additional interaction terms beyond the on-site $f-f$ repulsion included in \eqref{eq:action}, such as $c-f$ density-density and exchange interactions and $c-c$ interaction. 
All of these can be included perturbatively within our treatment. 
Within this work we will treat $c-f$ and $c-c$ density-density interactions within the Hartree approximation. 

The \textit{flat-chiral} limit of the THF model is given by taking $v_\star'=M=0$.
In this limit we can express the single-particle Hamiltonian as 
\begin{equation}
   \hat{h}^{(\lambda)}_\k =\left( \begin{array}{cc|c}
        0_{2\times2} &  v_\star\sigma_x(\k^{(\lambda)}\cdot\vec{\sigma}) & \gamma\sigma_0\\
        v_\star(\k^{(\lambda)}\cdot\vec{\sigma})\sigma_x & 0_{2\times2} & 0_{2\times2}
        \\\hline
         \gamma\sigma_0& 0_{2\times2}& 0_{2\times2}
    \end{array}\right).
\end{equation}
\vspace{0mm}

Due to its commutation with $I_{3\times3}\otimes\sigma_z$, $\hat{h}^{(\lambda)}_\k$ can be decoupled into two independent copies of a single $f$-electron coupled to a $c$-electron with Dirac dispersion. The two copies, corresponding to $\pm1$ eigenvalues of $\sigma_z$, have opposite chirality of the Dirac electrons. Altogether we have $N_f=4\times2$ such copies corresponding to the overall number of low-energy bands.

\vspace{5mm}

\section{Effective action for $c$ Electrons}
\label{section:expansion}
The effective action is derived by integrating out the $f$ electrons with respect to their bare action $S_f$. 
Before describing the derivation, we introduce some notations. 

Throughout the paper we use $\braket{}_0$ to denote correlators taken in the solvable theory given by $\hat{\gamma}_{ab}=0$. We use $\braket{}_c$ to denote connected correlators, and $\braket{}_{c,0}$ to denote connected correlators in the solvable theory.
We introduce a shorthand notation for local operators  $c_{i'}\equiv c_{\R'_i,\lambda'_i,a'_i}(\tau'_i)$, and $\bar{c}_i\equiv \bar{c}_{\R_i,\lambda_i,a_i}(\tau_i)$. Within connected correlators of the solvable theory, where all the $f$ and $\bar{f}$ must be of the same site, we use the shorthand notation $f_i\equiv f_{\R,\lambda_i,b_i}(\tau_i)$, $\bar{f}_{i'}\equiv \bar{f}_{\R,\lambda'_i,b'_i}(\tau'_i)$. For the hybridization we denote $\hat{\gamma}^\dagger_i\equiv \hat{\gamma}^\dagger_{a_ib_i}(\R-\R_i)$ and  $\hat{\gamma}_{i'}\equiv \hat{\gamma}_{b'_ia'_i}(\R-\R'_i)$. 

Following the strong coupling expansion~\cite{pairault1998strong,pairault2000strong} and the closely related cluster perturbation theory~\cite{Walenti1993,senechal2002cluster}, we trace out the $f$ electrons to derive an effective action for the $c$ electrons:
\begin{widetext}
\begin{equation}
    S_c^{\rm{eff}} = S_c-\log \braket{e^{-S_{fc}}}_0 = S_c - \sum_n \frac{1}{(n!)^2}\sum_{\substack{\{\lambda_i,a_i,\R_i\}\\\{\lambda'_i,a'_i,\R'_i\}}}
    \int_0^\beta \prod_{i=1}^n\left(d\tau_id\tau'_i\right)  \bar{c}_n \cdots\bar{c}_1
    \Gamma^{(n)}_{\substack{1\cdots n \\ 1'\cdots n'}}c_{1'}\cdots c_{n'} 
    ~,
    \label{eq:eff_action}
\end{equation}
with $\Gamma^{(n)}$ given by 
\begin{equation}
    \Gamma^{(n)}_{\substack{1\cdots n \\ 1'\cdots n'}} = \sum_\R\sum_{\{b_i\}\{b'_i\}}\prod_{i=1}^n (\hat{\gamma}^\dagger_i) \braket{f_1\cdots f_n \bar{f}_{n'} \cdots \bar{f}_{1'}}_{c,0}\prod_{i=1}^n  (\hat{\gamma}_{i'})~.
    \label{eq:Gamma_def}
\end{equation}
\end{widetext}
We note that this term is non-vanishing only if the set of tuples $\{(b_i,\lambda_i)\}$ equals the set $\{(b'_i,\lambda'_i)\}$. We also note that in the case of a momentum independent $\hat{\gamma}_\k$ we have $\hat{\gamma}(\R-\R')\propto\delta_{\R\R'}$, and thus all the vertices $\Gamma^{(n)}$ are local in space ($\R_i=\R'_j\equiv\R$), while in the general case they are non-local despite the locality of the $f$ correlators, due to the non-local nature of the $f-c$ hybridization.

To compute correlations within this theory we treat the quadratic terms $S_c$ and $\Gamma^{(1)}$ exactly, which reproduces the Hubbard I propagator (see Subsection~\ref{section:Hubbard_I} and Fig.~\ref{fig:header}c).
By treating all other $\Gamma^{(n)}$ interaction terms within a loop expansion we go beyond the Hubbard I approximation to calculate quantities such as the quasi-particle lifetime and flavor susceptibility within a strong coupling expansion, non-perturbatively in either $U$ or  $\hat{\gamma}_\k$. 

Crucially, we find that in the flat chiral limit, any diagram with $n_l$ loops contains a suppression factor of $s^{2n_l}$ (up to logarithmic factors in $s$). This result is derived in Appendix \ref{App:loop_counting}. The suppression of higher-loop diagrams drastically simplifies the theory, and allows for controlled calculations in many physical regimes of interest.
Away from the chiral limit, our explicit calculations show that the lowest-order loop corrections are still suppressed, but the dimensionless suppression factors have to be defined more carefully; see discussions below Equations~\eqref{eq:condition} and~\eqref{eq:excluded}.

\subsection{Deriving $f$ Correlators from a $c$ theory}
\label{section:f_corr}

\begin{figure}[t]
    \centering
\includegraphics[width=0.95\linewidth]{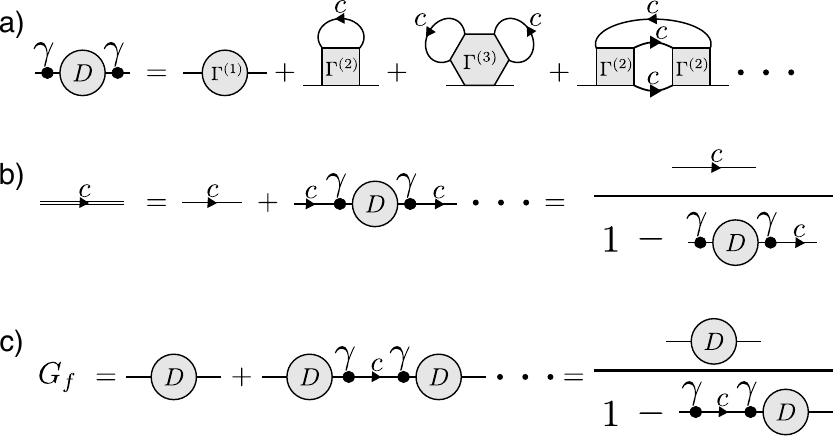}
    \caption{a) Leading contributions to the $c$ self-energy, also expressed as $\Sigma_c=\hat{\gamma}^\dagger D \hat{\gamma}$. b) The fully dressed $G_c$ expressed in terms of $\hat{\gamma}$, $D$ and $G^{0}_c$. c) $G_f$ computed in the fully-hybridized theory from the effective action containing only $c$-electrons. }
    \label{fig:f_c_relations}
\end{figure}

In deriving the expansion we integrated out the $f$-electrons.
However, we are often  interested in calculating $f$ correlators. These can be computed in the effective theory through exact relations between $f$ and $c$ correlators.

Let us begin by considering the two point function $G_{f}(k)=-\braket{f_k\bar{f}_k}$ with $k\equiv(\k,\omega)$, where we suppress flavor and orbital indices for simplicity. We define $D(k)$ as the irreducible part of $G_{f}(k)$ with respect to the bare propagator of the $c$-electrons, $G_{c}^{0}(k)=-\braket{c_k\bar{c}_k}_{0}$, such that  
\begin{align}
    G_{f} &=D+D\hat{\gamma}G_{c}^{0}\hat{\gamma}^\dagger D + \cdots \nonumber\\
    &= \frac{1}{D^{-1}-\hat{\gamma}G_{c}^{0}\hat{\gamma}^\dagger} .
    \label{eq:D_def}
\end{align}
In Fig.~\ref{fig:f_c_relations}.a we show examples of diagrams included in $D$.
Using this definition we can see that the full $c$ propagator $G_c(k)=-\braket{c_k\bar{c}_k}$ is given by
\begin{align}
    G_{c} &=G_{c}^{0}+G_{c}^{0}\hat{\gamma}^\dagger D\hat{\gamma}G_{c}^{0} + \cdots \nonumber\\
    &= \frac{1}{\left(G_{c}^{0}\right)^{-1}-\hat{\gamma}^\dagger D\hat{\gamma}},
    \label{eq:D_Gc_relation}
\end{align}
where by the definition of $D$ we avoid double-counting of diagrams.
This allows us to identify the self energy of the $c$-electrons with $D$ through the relation 
\begin{equation}
\label{eq:sigma_D_relation}
\Sigma_c(k)=\hat{\gamma}_\k^\dagger D(k)\hat{\gamma}_\k~.
\end{equation} 
Finally we note that for the case of the flat-chiral limit where the $c$-electrons are highly dispersive and the hybridization strength is momentum independent we can approximate Eq.~\eqref{eq:D_def} away from the Brillouin zone center by $G_f\approx D$. In the limit where $D$ is completely local (i.e. momentum independent) this gives a natural interpretation of $D$ as the local propagator of a single $f$ site, dressed by the $c$ electron bath.  

We define $\hat{\gamma}_\k^{-1}$ to be the right-inverse of $\hat{\gamma}_\k$ ($\hat{\gamma}_\k\hat{\gamma}_\k^{-1}=\mathbb{I}$) and by hermitian conjugation we define the left-inverse of  $\hat{\gamma}_\k^\dagger$ to express $G_f$ entirely in terms of the $c$ self-energy as
\begin{equation}
    G_{f} = \frac{(\hat{\gamma}^{-1})^\dagger\Sigma_c\hat{\gamma}^{-1}}{1-\hat{\gamma}G_{c}^{0}\Sigma_c\hat{\gamma}^{-1}}~.
    \label{eq:f_prop}
\end{equation}
Lastly, we can derive the equivalent reciprocal relation
\begin{equation}
    G_{c}-G_{c}^{0}=G_{c}^{0}\,\hat{\gamma}^\dagger G_{f}\,\hat{\gamma} \,G_{c}^{0}~, 
    \label{eq: f_2_point}
\end{equation}
as can be seen from the diagrammatic depiction of equations \eqref{eq:D_def} and \eqref{eq:D_Gc_relation} in Fig.~\ref{fig:f_c_relations}.c and Fig.~\ref{fig:f_c_relations}.b respectively.

This relation can be generalized to higher order cumulants. Since the $c$-electrons are non-interacting for $\gamma = 0$, any connected diagram beyond the two-point function vanishes, $\braket{cc \cdots \bar{c}\bar{c}}_{c,0}=0$. 
We thus see that beyond the two point function, Eq.~\eqref{eq: f_2_point}, generalizes to the simple form
\begin{multline}
    \braket{c_{k_1}c_{k_2} \cdots \bar{c}_{k'_2}\bar{c}_{k'_1}}_c= \\ \prod_i\left(G_{c}^{0}(k_i)\hat{\gamma}_\k^\dagger\right)\braket{f_{k_1}f_{k_2}\cdots \bar{f}_{k'_2}\bar{f}_{k'_1}}_c\prod_i \left( \hat{\gamma}_\k G_{c}^{0}(k'_i)\right)~.
\end{multline}
This identity also applies to mixed correlators, where $G_{c}^{0}(k) \hat{\gamma}_\k^\dagger$ takes $f_k\rightarrow c_k$, and by conjugation $\hat{\gamma}_\k G_{c}^{0}(k)$ takes $\bar{f}_k\rightarrow\bar{c}_k$.
For example we can write
\begin{equation}
    \braket{c_{k_1}c_{k_2}\cdots \bar{c}_{k'_2}\bar{c}_{k'_1}}_c= G_{c}^{0}(k_1) \hat{\gamma}^\dagger \braket{f_{k_1}c_{k_2}\cdots \bar{c}_{k'_2}\bar{c}_{k'_1}}_c~.
\end{equation}
To reinstate flavor and orbital indices, we define $G_{c}^{0}(k,\lambda;a,a')=\braket{c_{k,\lambda,a}\bar{c}_{k,\lambda,a'}}_0$
to have $G_{c}^{0}(k,\lambda;a,a') (\hat{\gamma}_\k^\dagger)_{a'b}$ taking $f_{k,\lambda,b}\rightarrow c_{k,\lambda,a}$ with $a'$ being summed over using the Einstein summation convention.
We emphasize the fact that all of the derived relations in this section rely solely on the absence of interactions between $c$-electrons and between $c$ and $f$ electrons, and are thus exact within our model.

\section{Quasiparticle Self Energy}
\label{section:QP}

\subsection{Hubbard I approximation: Tree-Level Self Energy}
\label{section:Hubbard_I}

In Sec.~\ref{section:f_corr} we identified the self energy of the $c$ electrons with $D$, the irreducible part of the $f$ two point function through the relation $\hat{\Sigma}_c = \hat{\gamma}^\dagger D\hat{\gamma}$, derived from Eq.~\eqref{eq:D_def},\eqref{eq:D_Gc_relation}. To tree-level  order there is only one contribution to $D$ given by the non-hybridized single-body $f$ propagator
\begin{align}
    D^{(0)}(i\omega) &=\Gamma^{(1)}=G_f^0(i\omega)\nonumber\\
    &=\frac{\frac{1}{2}-\frac{\nu}{N_f}}{i\omega-E_+}+\frac{\frac{1}{2}+\frac{\nu}{N_f}}{i\omega+E_-} 
    \overset{\mu=0}{\rightarrow}\frac{-i\omega}{\omega^2+u^2}~,
\end{align} 
where we define $\nu = \braket{\delta n_\R}_0$, the deviation from half-filling, $E_\pm$ is the excitation energy of either the upper or the lower Mott band of the non-hybridized theory given by $E_\pm = \left|\pm u-\delta\mu \right|$, with $u=\frac{U}{2}$ and $\delta\mu= \mu-\nu U$. Notice that $\nu$ is always an integer, because $\delta n_\R$ is a good quantum number in the $\hat{\gamma}_\k=0$ theory and we assume that $T\ll U$. Thus, the thermal expectation value in the single site problem is given by its value in the ground-state manifold.

To tree-level order we have
\begin{equation}
    G_{c,\lambda}(\k,i\omega) = \frac{1}{i\omega-\hat{h}_\k^{(c,\lambda)}-\hat{\gamma}_\k^\dagger G_f^0(i\omega)\hat{\gamma}_\k}.
    \label{eq:Gc_H1}
\end{equation}
Similarly we can find $G_f$ using eq.~\eqref{eq:D_def}. This result is known as the Hubbard I propagator~\cite{hubbard1963electron}. 
In real frequencies ($i\omega\rightarrow\omega+i0^+$) we find the poles of Eq.~\eqref{eq:Gc_H1} at charge neutrality, and in the flat-chiral limit, to be given by
\begin{equation}
    v_\star^2|\k|^2=\omega^2\left(1+\frac{\gamma^2}{u^2-\omega^2}\right).
\end{equation}
This equation describes the dispersion of a ``Mott semimetal,'' with flat Hubbard-like bands at $\omega \approx \pm u$ for $v_\star|\k|/\gamma \gg 1$ and a gap closing at $\k=0$. In the band-projected limit ($\gamma \gg u$) it reduces to \begin{equation}
    \frac{v_\star}{\gamma}|\k|=\pm\frac{\omega}{\sqrt{u^2-\omega^2}},
\end{equation}
which coincides with the result of Ref.~\cite{ledwith2024nonlocal} upon identifying 
\begin{equation}
    s^2\equiv\frac{\gamma^2}{2v_\star^2}\frac{2\pi}{A_{BZ}},
    \label{eq:s2}
\end{equation} as first noted in Ref.~\cite{hu2025projected}. The poles of $G_f$ coincide with those of $G_c$, whereas the corresponding residues, $Z_c$ and $Z_f$, encode the orbital content of the quasiparticles. 

Ref.~\cite{ledwith2024nonlocal} showed that in the projected limit, adding a non-zero single-particle bandwidth ($M\ne0$) results in a vanishing quasiparticle residue for the remaining pole at $\omega=\k=0$. Let us now consider the same scenario away from the projected limit. Using Eq.~\eqref{eq:D_def},\eqref{eq:D_Gc_relation} we find
\begin{align}
    G_f(\k=0,i\omega) &= \frac{1}{\frac{\omega^2+u^2}{-i\omega}-\frac{\gamma^2}{i\omega}}\nonumber\\
    &= \frac{1/2}{i\omega+\sqrt{\gamma^2 +u^2}}+\frac{1/2}{i\omega-\sqrt{\gamma^2 +u^2}}~,\\
    G_c(\k=0,i\omega) &= \frac{1}{i\omega-M\sigma_x\left(\frac{\zeta_0-\zeta_z}{2}\right)+i\omega\frac{\gamma^2}{\omega^2+u^2}\left(\frac{\zeta_0+\zeta_z}{2}\right)} \nonumber\\
    &= \frac{\frac{\zeta_0-\zeta_z}{2}}{i\omega-M\sigma_x} + \frac{\frac{\zeta_0+\zeta_z}{2}}{i\omega(1+\frac{\gamma^2}{\omega^2+u^2})}~.
\end{align}
The zero-energy pole in the $c$ propagator has a finite residue 
\begin{equation}
    Z_{c,\zeta_z=1}(\k=0,\omega=0) =\frac{u^2}{u^2+\gamma^2},
    \label{eq:c_res}
\end{equation} 
algebraically suppressed in the large $\gamma/u$ limit. The residues for $f$ electrons and $c$ electrons with $\zeta=-1$ vanish at $\omega=0$. This can be understood by the symmetry properties of the excitations at the $\Gamma$ point. In the band projected limit, this excitation has a large overlap with an $f$ electron trion operator, composed of an $f$ electron/hole times an electron-hole pair~\cite{ledwith2025exotic}.

\subsection{$f$ Four Point Function}
\label{section:4f}

In order to go beyond the Hubbard I approximation, we calculate the four $f$ cumulant that appears in $\Gamma^{(2)}$:
\begin{multline}
     \braket{f_1\left(\tau_{1}\right)f_2\left(\tau_{2}\right)\bar{f}_2\left(\tau'_{2}\right)\bar{f}_1\left(\tau'_{1}\right)}_{c,0}\equiv\\
    \braket{f_1f_2\bar{f}_2\bar{f}_1}_0
    -\braket{f_1 \bar{f}_1}_0\braket{f_2\bar{f}_2}_0 + \braket{f_2 \bar{f}_1}_0 \braket{f_1\bar{f}_2}_0.
\end{multline}
We implicitly take all $f$ and $\bar{f}$ to be at the same position $\R_i$, as the cumulant vanishes otherwise. We hereby focus on the long-time behavior leading to a finite scattering rate at one-loop order, with the full expression given in App.~\ref{App:f4}.
The dominant contribution to this cumulant at long imaginary-time differences, $\left|\tau_i-\tau'_i\right|\gg\frac{1}{\min (E_+,E_-)}$, can be interpreted as a flavor flip process (see Fig.~\ref{fig:header}.b), with a flavor flip a time $\tau_1\sim\tau'_2$ and another one at time $\tau_2\sim\tau'_1$ 
. For this correlator to be of appreciable value we need to satisfy $\left|\tau_i-\tau'_j\right|\lesssim\frac{1}{\min (E_+,E_-)}$ for $j\ne i$. 
The intermediate state between the flavor-flips, at times $\min(\tau_1,\tau_2)\ll \tau\ll \max(\tau_1,\tau_2)$, is a ground state of the single-site problem described by $S_f$, as it has the same charge as the initial (ground-)state and $S_f$ depends only on the total charge. For this reason, this correlator is not suppressed in the time difference $\max(\tau_1,\tau_2)-\min(\tau_1,\tau_2)$, taken here to be long. 

In the long-time difference limit, and for different flavor-orbital combinations $(b_1,\lambda_1) \ne(b_2,\lambda_2)$, the disconnected contribution is negligible, and we have
\begin{multline}
    \braket{f_1\left(\tau_{1}\right)f_2\left(\tau_{2}\right)\bar{f}_2\left(\tau'_{2}\right)\bar{f}_1\left(\tau'_{1}\right)}_{c,0} \approx -\frac{1-(2\nu/N_f)^2}{4(1-N_f^{-1})}\\ 
    \times\left[ \Theta(\tau_2-\tau'_1)e^{-E_+(\tau_2-\tau'_1)} + \Theta(\tau'_1-\tau_2)e^{-E_-(\tau'_1-\tau_2)}\right]\\
    \times\left[ \Theta(\tau_1-\tau'_2)e^{-E_+(\tau_1-\tau'_2)} + \Theta(\tau'_2-\tau_1)e^{-E_-(\tau'_2-\tau_1)}\right].
    \label{eq:4pt_tau}
\end{multline}
The filling dependent pre-factor accounts for the probability to finds a state with $(b_1,\lambda_1)$ occupied and $(b_2,\lambda_2)$ empty or vice versa. The case of $(b_1,\lambda_1)=(b_2,\lambda_2)$ is different due to the non-vanishing disconnected part (see App.~\ref{App:f4} for details). We will neglect contribution of the same flavor-orbital type, as they are subleading in $N_f^{-1}$, which for MATBG is small ($N_f=8$). 

Next, we take the physical limit of no thermal charge fluctuations ($\beta E_\pm \gg 1$) to neglect temporal overlap between the two flavor-flip instantons. We find the Fourier transform of this contribution to be
\begin{align}
    \label{eq:f_4_point}
    \left\langle f_{1,i\omega_{1}}f_{2,i\omega_{2}}\bar{f}_{2,i\omega'_{2}}\bar{f}_{1,i\omega'_{1}}\right\rangle_{c,0} &= -\frac{1-(2\nu/N_f)^2}{4(1-N_f^{-1})} \nonumber\\
    \times\delta_{\omega'_{2},\omega_{1}}\delta_{\omega'_{1},\omega_{2}}
    \prod_{j=1,2}&\left(\frac{1}{E_+-i\omega_{j}} + \frac{1}{E_-+i\omega_{j}} \right)\nonumber\\
    +&\text{(additional terms),}
\end{align}
where we have used the shorthand notation $f_{i,\omega_i} \equiv f_{\R,\lambda_i,b_i,\omega_i}$ and 
$\bar{f}_{i,\omega'_i} \equiv \bar{f}_{\R,\lambda_i,b_i,\omega'_i}$.

In addition to this contribution, the four-point cumulant includes short range in time terms, where all four operators are within an imaginary time interval of the order of $\frac{1}{\min (E_+,E_-)}$,
and a contribution with the opposite limits of long and short time differences (i.e. $\left|\tau_i-\tau'_i\right|\ll\frac{1}{\min (E_+,E_-)}$ and $\left|\tau_1-\tau_2'\right|\gg\frac{1}{\min (E_+,E_-)}$).
The short-range in time contributions cannot lead to a finite scattering rate at low energies at one-loop order. This follows from the optical theorem, since the one-loop diagram cannot be cut in a way that corresponds to an on-shell scattering process. 

The additional extended-in-time contribution, having $\left|\tau_1-\tau'_2\right|\gg\frac{1}{\min (E_+,E_-)}$, cancels with the disconnected correlator up to order $N_f^{-1}$ and can be neglected in the large $N_f$ limit. 
In fact, at one loop order one can show that the contribution from this term vanishes exactly for any finite $N_f$ due to the fact that in this limit of $\left|\tau_i-\tau_j'\right|\gg\frac{1}{\min (E_+,E_-)}$ (with $i\ne j$) we have\footnote{The difference between the full and the disconnected correlator in this limit can be fully attributed to conditional probability of finding $(b_i,\lambda_i)$ occupied (unoccupied) given that $(b_j,\lambda_j)$ is occupied (unoccupied). However, the total number of occupied (unoccupied) $f$ electrons is predetermined as a property of the ground-state manifold.  Therefore, upon summation over all flavor-orbital combinations, the connected and disconnected contributions exactly cancel.}
\begin{equation}
    \sum_{(b_2,\lambda_2)} \braket{f_1\left(\tau_{1}\right)f_2\left(\tau_{2}\right)\bar{f}_2\left(\tau'_{2}\right)\bar{f}_1\left(\tau'_{1}\right)}_{c,0} \rightarrow 0.
\end{equation}

Following these arguments we will approximate $\Gamma^{(2)}$ using the expression in Eq.~\eqref{eq:f_4_point}, neglecting other contributions to $\braket{f_1f_2\bar{f}_2\bar{f}_1}$. Later in the text we take a similar approach to analyze contribution from $\Gamma^{(n)}$ with $n>2$ by considering similar multiple flavor-flips processes.

\subsection{One Loop Order}
\label{section:one loop}
To one loop order, the only diagram that contributes to $D$ is a contraction of the two-body vertex 
\begin{equation}
    \hat{\gamma}_1^\dagger D^{(1)}(\omega)\hat{\gamma}_1 = \sum_{\lambda'}\int_{\omega',\k}
    \Gamma^{(2)}_{\lambda,\lambda'}(\omega,\omega';\omega',\omega)G_{c,\lambda'}(\k,\omega')~,
    \label{eq:SE}
\end{equation}
where, $\int_{\omega'}\equiv\int\frac{d\omega'}{2\pi}$, $\int_{\k}\equiv\int\frac{d^2\k}{A_{BZ}}$ and we suppressed orbital indices for brevity. $\lambda$ on the RHS is the flavor index of $D^{(1)}$. It is omitted from the LHS because $D^{(1)}$ is $\rm{SU}\left(N_\lambda\right)$ symmetric.

We define the hybridization integral as the fermionic loop integral at frequency $\omega$ weighted by two $\hat{\gamma}_\k$ factors:
\begin{equation}
    \mathcal{I}(i\omega) = \sum_\lambda\int \frac{d^2k}{A_{BZ}}
    \operatorname{Tr}\left[
    \hat{\gamma}_\k G_{c,\lambda}(\k,i\omega) \hat{\gamma}_\k^\dagger\right].
    \label{eq:hybrid_int}
\end{equation}
Together with the expression for $\Gamma^{(2)}$ in Eq.~\eqref{eq:f_4_point} we arrive at the expression for $D$ to one-loop order
\begin{align}
    D^{(1)}(i\omega)= \left[ 1-\left(\frac{2\nu}{N_f}\right)^2 \right]
     \frac{u^2\mathcal{I}(i\omega) }{\left[u^2-(i\omega+\delta\mu)^2\right]^2}  
    ~,
    \label{eq:D1}
\end{align} 
up to a $N_f^{-1}$ correction from the term where the flavor in the loop equals the external one.

For the dispersive part of the spectrum, sufficiently far from the Mott bands, we can demand the one-loop correction to be subleading to the tree-level contribution
\begin{equation}
    \left|\frac{D^{(1)}}{D^{(0)}}\right|\sim\left|\frac{\mathcal{I}/u^2}{\omega/u^2}\right|=\frac{|\mathcal{I}|}{|\omega|}\ll1.
    \label{eq:condition}
\end{equation}
We thus find $\mathcal{I}/u$ to be a natural small parameter to control the expansion.
To connect between this small parameter and $s^2$, the small parameter in the chiral limit, we calculate $\mathcal{I}$ at charge neutrality and in the flat-chiral limit. We carry the calculation to tree-level approximation:
\begin{align}
    \mathcal{I}(i\omega)&=N_f\int \frac{d^2k}{A_{BZ}}\frac{-i\omega\gamma^2}{\omega^2(1+\frac{\gamma^2}{\omega^2+u^2})+v_\star^2|\k|^2}\nonumber\\
    &\approx -i\omega\gamma^2N_f\frac{2\pi}{A_{BZ}} \int_0^{\sqrt{\frac{A_{BZ}}{\pi}}}\frac{kdk}{\omega^2(1+\frac{\gamma^2}{\omega^2+u^2})+v_\star^2k^2} \nonumber\\
    &= -i\omega N_f\frac{\gamma^2}{2v^2_\star}\frac{2\pi}{A_{BZ}} \log \frac{\omega^2(1+\frac{\gamma^2}{\omega^2+u^2})+\frac{v_\star^2A_{BZ}}{\pi}}{\omega^2(1+\frac{\gamma^2}{\omega^2+u^2})} \nonumber\\
    \overset{\gamma\gg\omega, u}{\Rightarrow}&= -i\omega N_f\frac{\gamma^2}{2v^2_\star}\frac{2\pi}{A_{BZ}} \log \frac{\gamma^2+\frac{v_\star^2A_{BZ}}{\pi}\frac{\omega^2+u^2}{\omega^2}}{\gamma^2} \nonumber\\
    &=-i\omega N_f s^2 \log\left(1+\frac{1}{s^2}\frac{\omega^2+u^2}{\omega^2}\right).
    \label{eq:I}
\end{align}
The condition $\left| \mathcal{I}(i\omega) \right|\ll u$ for $\omega\lesssim u$ therefore corresponds to $N_f s^2\log\left( \frac{2}{s^2} \right) \ll 1$. In fact that is the correct small parameter in the chiral limit. It includes a factor of $N_f$, the number of $f$ flavor-orbitals, as arises from the explicit calculation of the one-loop self energy.

Unlike $D^{(0)}$, the one loop correction $D^{(1)}$, and therefore the one-loop self-energy, has an imaginary part for any real frequency within the low-energy bands. This gives the quasiparticles a finite lifetime. 
At charge neutrality, taking the flat-chiral limit, we use Eq.~\eqref{eq:gamma_def}, \eqref{eq:hybrid_int},\eqref{eq:sigma_D_relation} and~\eqref{eq:D1} to find the imaginary part of the real-frequency self-energy:
\begin{equation}
    \Sigma''_c(\omega\ll u)= -\left( \frac{\zeta_0+\zeta_z}{2}\right)
     \frac{\gamma^4}{u^2}\pi\rho_{c,\zeta_z=1}(\omega),
\end{equation}
with
\begin{align}
    \rho_{c,\zeta_z=1}(\omega) =-\frac{1}{\pi}\int_\k\operatorname{Im}\operatorname{Tr}\left[ \frac{\zeta_0+\zeta_z}{2}\hat{G}_c(\k,\omega)\right]
    \label{eq:rho_c}
\end{align} 
the density of states of $c$-electrons with orbital character $\zeta_z=+1$. This result has a simple and direct interpretation: it represents elastic scattering of  $\zeta_z=+1$ $c$ electrons off the $f$ local moments, with $\frac{\gamma^2}{u}$ being the scattering matrix element. In this low energy limit the $f$ electron moments are effectively quasi-static. 

To correctly compute the quasiparticle lifetime we need to account for its residue. we find the inverse lifetime for a quasiparticle to be
\begin{align}
\label{eq:dispersive_lifetime}
    \tau^{-1} (\k,\omega\ll u) &= -Z_{c,\zeta_z=1}(\k,\omega)  \Sigma''_{c,\zeta_z=1}(\omega) \nonumber\\
    &=\pi\frac{\gamma^4}{u^2}Z_{c,\zeta_z=1}(\k,\omega)\rho_{c,\zeta_z=1}(\omega). 
\end{align}
In the chiral limit,  $\operatorname{Im}\mathcal{I}(\omega+i0^+)=-\pi\gamma^2\rho_{c,\zeta_z=1}(\omega)$. Using this relation together with the expressions for $\mathcal{I}$ in Eq.~\eqref{eq:I} and $Z_{c,\zeta_z=1}$ in Eq.~\eqref{eq:c_res}, we find
\begin{equation}
 \tau^{-1}(\k,\omega\ll u)  
 = \frac{\pi|\omega|N_f s^2}{1 + (u/\gamma)^2}.    
\end{equation}
Note that while the residue of the $\zeta=+1$ $c$-electron at the $\Gamma$ point asymptotically vanishes in the projected limit, its scattering still gives an overall finite quasiparticle scattering rate, as the vanishing residue is counteracted by the diverging scattering matrix element $\tfrac{\gamma^2}{u}$.

We note that away from the flat limit ($M\ne0$) the gapped quasi-particles at $\k=0$, $\omega=\pm M$ are of orbital character $\zeta_z=-1$, thus having $Z_{c,\zeta_z=1}=0$. Within the model considered, this will result in infinitely long lived gapped quasiparticles at the $\Gamma$ point. This is an artifact of our model. A symmetry allowed $f-c$ exchange interaction (denote by $J$ in App.~\ref{App:THF_model}) will give these quasi-particles a finite lifetime at one loop order through a similar elastic scattering with scattering matrix element $J$ instead of $\tfrac{\gamma^2}{u}$.

\subsection{Infinite Loop resummation: Resonance Broadening in the Large $N_f$ Limit}
\label{section:loop-resummation}
As we cross over from the dispersive to the Mott-like part of the band ($\omega\rightarrow\pm E_\pm$), the one-loop self-energy diverges [see Eq.~\eqref{eq:D1}]. 
The divergence originates from the fact that $\Gamma^{(2)}$, the effective interaction, is derived from the bare  $f$ single-site theory ($\hat{\gamma}=0$), having a sharp discrete spectrum. This is analogous to the standard divergence in weak-coupling perturbation theory for the propagator, resolved by a Dyson resummation. In our case, due to the integration out of the $f$ degrees of freedom, a resummation of one-particle-irreducible (1PI) diagrams does not cure all divergences. We need to identify an infinite set of the most divergent diagrams and resum them explicitly.

In order to identify the relevant set of diagrams we take the frequency deviation from the Mott band to be of the order of the small hybridization integral $\omega\mp E_\pm\sim\mathcal{I}(\omega)$. In this regime we need to keep all contributions of the form $ \tfrac{\mathcal{I}(\omega)^n}{\left(\omega\mp E_\pm\right)^{n+1}}\sim \tfrac{1}{\omega\mp E_\pm}$ as they are comparable to the tree-level self energy.

\begin{figure}
    \centering
    \includegraphics[width=.95\linewidth]{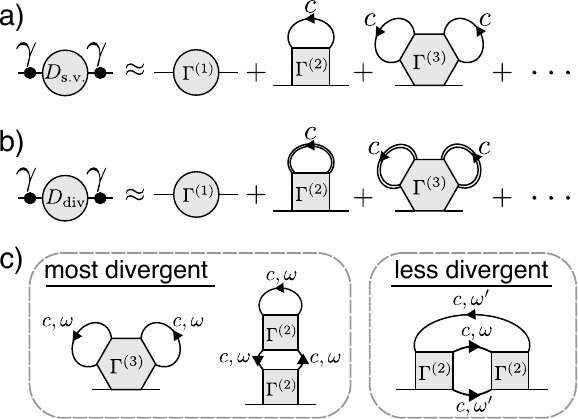}
    \caption{a) The set of diagrams accounted for in the `single-vertex' self-energy $\hat{\gamma}^\dagger D_\text{s.v.}\hat{\gamma}$. The approximations made are accounting only for the leading term in $\frac{1}{N_f}$, and only for the contribution which is most divergent in $\frac{\mathcal{I}_0}{\omega\mp E_\pm}$. b) The set of diagrams accounted for in the $\hat{\gamma}^\dagger D_\text{div}\hat{\gamma}$. The approximations made are the same as in a. This expression accounts for all of the most diverging contributions at any order in $\mathcal{I}_0$ c) 
    All of the diagrams at two-loops order. The first two has a most divergent part to them, with all propagators taken at the external frequency $\omega$. The right-most diagram does not have a contribution of the most divergent form, all parts of the $\Gamma^{(2)}$ interaction result in at least two propagators taken at an independent, integrated over, frequency $\omega'$. The frequencies that are not integrated over are a result of the delta functions in the long-range part of $\Gamma^{(n)}$ (see for example Eq.~\eqref{eq:f_4_point}).}
    \label{fig:SE_resummation}
\end{figure}

Consider the contribution to the $c$ self-energy from the diagram composed of a single loop-contracted $\Gamma^{(n+1)}$ vertex (Fig.~\ref{fig:SE_resummation}.a). 
For certain time orderings, the contribution from this diagram in imaginary-time is given by a chain of flavor-flips separated by long time intervals, analogous to the contribution we found at one loop (Sec.~\ref{section:one loop}). In frequency domain, as we later show, this results in a contribution where $c$ electron loops, and all flavor-flip instantons, are at the external frequency. Each flavor-flip `instanton' contributes a factor of $\tfrac{1}{\omega\mp E_\pm}$ such that the overall contribution of this multi flavor-flip process is proportional to $\tfrac{\mathcal{I}(\omega)^n}{\left(\omega\mp E_\pm\right)^{n+1}}$ and therefore must be accounted for in the resummation. In addition, any insertion of such self-energy diagram along one of the $c$ loops  will give an additional factor of $\left(\tfrac{\mathcal{I}(\omega)}{\omega\mp E_\pm}\right)^{m}\sim1$. Hence the loops must be computed using the dressed $c$ propagators, in a self-consistent manner (Fig.~\ref{fig:SE_resummation}.b).

Lastly, we argue that diagrams that contain more than a single vertex (For example the right hand part of Fig.~\ref{fig:SE_resummation}.c) can be safely neglected in this regime. The argument for this is that such diagram always contain internal frequencies that are integrated over, such that not all instantons are at the external frequency and therefore the power of the divergent term for a given number of loops is smaller. In the chiral limit, where the number of loops directly correspond to the power of $s^2$, this immediately means that these diagrams are sub-leading. Away from the chiral limit, we argue that the same holds given that the appropriate parameter is small, we calculate explicitly one such diagram in Sec.~\ref{section:excluded}.

We proceed to sum the relevant diagrams. For analytical tractability, we will take the limit of $N_f\rightarrow\infty$, where we can safely assume that all the $f$ flavor-orbitals entering the multipoint correlator are distinct.\footnote{Formally, we scale $v_\star^2$ with $N_f$, in order to keep the small parameter $\mathcal{I}/u\sim s^2 N_f$ fixed.} We first solve for charge-neutrality ($\nu=0$) and then generalize to $\nu \ne 0$.

In the technical part below we establish the following: 1) In the limit $\beta E_\pm,N_f\gg1$, the dominant contributions to $D$ from a single $\Gamma^{(m)}$ vertex insertion (see Fig.~\ref{fig:SE_resummation}.b) is described by multi-instanton (flavor-flip) processes.
2) These instantons admit a canonical ordering according to the flavors of the created and annihilated $f$ electrons, independently of their time ordering. 
3) Within this canonical ordering, excitations of the $c$ electron bath (denoted here by $\tilde{\mathcal{I}}(\tau)$) connect subsequent instantons such that the multi-instanton contribution can be written as a convolution in time domain.  
4) This can be used to derive a simple expression for the divergent contribution of each diagram in Fig.~\ref{fig:SE_resummation}.b, and sum all of these contributions (which form a geometric series) to arrive at the expression in Eq.~\eqref{eq:most_div_resummed}.  

\begin{widetext}
We begin by considering the $m-1$ loop diagram, contracting a single $\Gamma^{(m)}$ vertex, given by 

\begin{align}
    &\sum_{\R,\R'}\int_{\tau,\tau'}\tilde{G}_c^0(x'_0-x) \tilde{\Sigma}_{m-1}(x-x')\tilde{G}_c^0(x'-x_0) \nonumber\\
    &= \frac{1}{(m!)^2}\int_{\{\tau_i\},\{\tau'_i\}}\sum_{\substack{\{\lambda_i,a_i,\R_i\}\\\{\lambda'_i,a'_i,\R'_i\}}} \Gamma^{(m)}_{\substack{1\cdots m \\ 1'\cdots m'}}\left[\braket{\bar{c}_0c_{0'}\bar{c}_m \cdots\bar{c}_1 c_{1'}\cdots c_{m'} }-\braket{c_{0'}\bar{c}_0}\braket{\bar{c}_m \cdots\bar{c}_1 c_{1'}\cdots c_{m'}}\right]\nonumber\\
\end{align}
Under the assumption of large $N_f$ (see App.~\ref{App:Large_N} for details), it  reduces to
\begin{equation}
\label{eq:single_vertex_contr}
    \tilde{D}_{m-1}(\tau_m-\tau'_m)\approx -\frac{1}{(m-1)!}\int \prod_{i=1}^{m-1}\left(d\tau_i d\tau_i'\right)  \braket{f_1\cdots f_m \bar{f}_{m'} \cdots \bar{f}_{1'}}_{c,0} \prod_{i=1}^{m-1}\tilde{\mathcal{I}}(\tau'_i-\tau_i),
\end{equation}
\end{widetext}
where here $f_i\equiv f_{\R,\lambda_i,b_i}(\tau_i)$ and $\bar{f}_{i'}\equiv \bar{f}_{\R,\lambda_i,b_i}(\tau'_i)$ have the same flavor and orbital.

\begin{figure}
    \centering
    \includegraphics[width=0.9\linewidth]{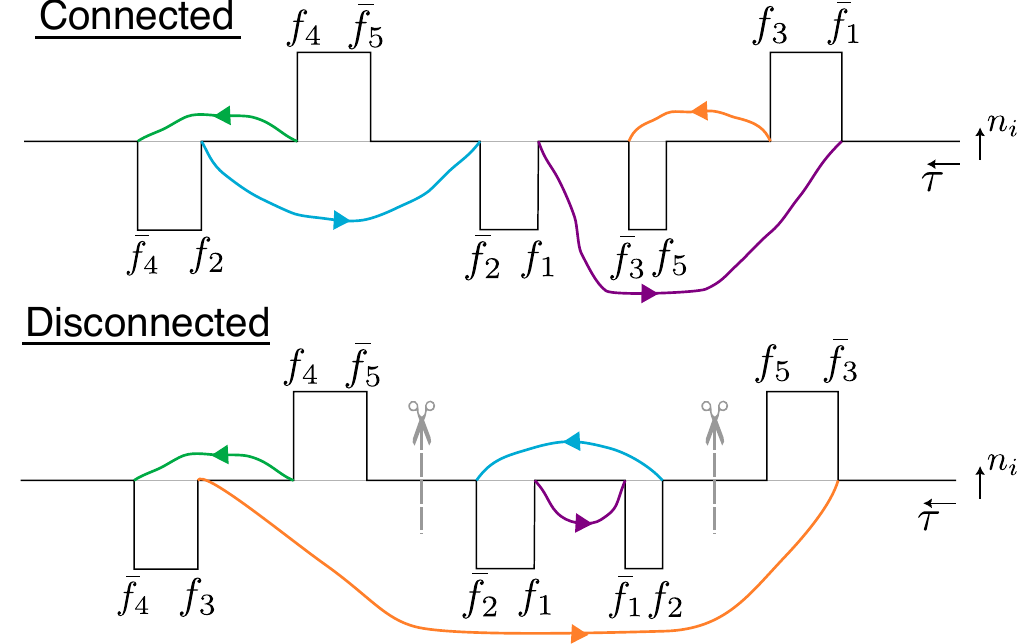}
    \caption{Graphic representation of two different five-instanton processes. The height of the black line depicts the charge of the $f$ site at each point in time, with the equilibrium (ground-state) charge marked by a gray line. The colorful lines represent excitations of the $c$-electron bath (the fermionic loops in Fig. \ref{fig:SE_resummation}) with the color representing the flavor of the excitation. The instantons (doublons/holons) are the states where the charge of the site deviates from its equilibrium value and are confined in time by the charging energy. 
    For the top panel, the mapping from the created to the annihilated flavor at each instanton is cyclic. The flavor occupation of the site in this configuration does not repeat itself at any intermediate time. 
    In contrast, the mapping in the bottom panel includes the sub-cycle $1\rightarrow2\rightarrow1$. The connected $f$ correlator in this case vanishes to leading order in $1/N_f$, since the full correlator cancels with the disconnected term
    $\langle f_1 f_2 \bar{f}_2 \bar{f}_1\rangle \langle f_3 f_4 f_5 \bar{f}_5 \bar{f}_4 \bar{f}_3 \rangle$,
    as indicated by the dashed gray lines.
    }
    \label{fig:cyclic contributions}
\end{figure}

Next we define the permutation maps $K:i\rightarrow k_i$ and $L:i\rightarrow\ell_i$ as the time ordering permutations for creation and annihilation operators separately such that $\tau_{k_1}<\tau_{k_2}<\cdots\tau_{k_m}$ and $\tau'_{\ell_1}<\tau'_{\ell_2}<\cdots\tau'_{\ell_m}$. For time orderings that satisfy $\tau'_{\ell_i}<\tau_{k_{i+1}}$ and $\tau_{k_i}<\tau'_{\ell_{i+1}}$ the $f$ correlator is described by a set of non-overlapping ``instantons'', where each instanton corresponds to a charge $+1$ (``doublon'') or $-1$ (``holon'') excitation of the unhybridized $f$ site (see Fig. \ref{fig:cyclic contributions}). 

In the non-overlapping instanton case, specializing to half-filling, we can express 
\begin{align}
    \label{eq:f_multipoint}
    &\braket{f_1\cdots f_m \bar{f}_{m'} \cdots \bar{f}_{1'}}_{0} 
    \nonumber\\
    \approx &\sgn_\tau\prod_{i=1}^m -\frac{1}{2}\left[\Theta\left(\delta\tau_{i}\right)e^{-E_+\delta\tau_{i}} + \Theta\left(-\delta\tau_{i}\right)e^{-E_-\left(-\delta\tau_{i}\right)} \right]\nonumber\\
    \equiv&\sgn_\tau\prod_{i=1}^m \tilde{G}_{ins}(\delta\tau_i=\tau_{k_i}-\tau_{\ell_i}')~,
\end{align}
where the last line defines the instanton Green's function, $\tilde{G}_{ins}(\tau)$. The global sign from time ordering is given by  $\sgn_\tau=\prod_i \sgn(\tau'_{i}-\tau_{i})$ (see App.~\ref{App:time_ordering}).
The combinatorial prefactor of $\frac{1}{2}$ per instanton is correct to leading order in $N_f^{-1}$, consistent with the large $N_f$ limit. 

Eq.~\eqref{eq:f_multipoint} is the expression for the full correlator, whereas Eq.~\eqref{eq:single_vertex_contr} contains only the connected part. 
In the dilute instanton gas limit we have two possible cases. In the first case the disconnected part is exponentially suppressed when the inter-instanton time separation is larger than $1/u$, in which case we approximate the connected correlator by the full correlator. In the second case, the disconnected part is equal to the full correlator up to a correction of order $N_f^{-1}$, in which case the connected correlator vanishes to leading order $N_f^{-1}$.

The condition for the connected correlator to be of the non-vanishing type is formulated as follows. 
We consider the permutation map $P=L\circ K^{-1}:k_i\rightarrow i\rightarrow \ell_i$ which maps between the flavor that is created to the flavor that is annihilated at each instanton. To get a disconnected contribution which is not exponentially suppressed relative to the full correlator we must take expectation values over sub-sets of operators that create complete cycles in $P$.
Therefore, if $P$ does not contain sub-cycles (i.e. $P$ is cyclic) all disconnected correlators are suppressed and we have $\braket{\cdots}_{c,0}\approx \braket{\cdots}_{0}$. Otherwise, to leading order in $N_f^{-1}$, the contribution cancels with the disconnected part. Examples of contributions of both types are shown in Fig.~\ref{fig:cyclic contributions}.  

Finally, for time orderings that satisfy the cyclic constraint, we can write the integrand in Eq. \eqref{eq:single_vertex_contr} as

\begin{align}
    \label{eq:convolution_form}
    &\;\;\;\sgn_\tau\prod_{i=1}^m \tilde{G}_{ins}(\tau_{k_i}-\tau_{\ell_i}') \prod_{i=1}^{m-1} \tilde{\mathcal{I}}(\tau'_i-\tau_i) \nonumber \\
    &
    \begin{aligned}
        =\sgn(\tau'_m-\tau_m)&\prod_{i=1}^m \tilde{G}_{ins}(\tau_{k_i}-\tau_{\ell_i}') \\ \times&\prod_{i=1}^{m-1} \sgn(\tau'_i-\tau_i)\tilde{\mathcal{I}}(\tau'_i-\tau_i)
    \end{aligned}
     \nonumber \\
    &
    \begin{aligned}
        =\sgn(\tau'_m-\tau_m)&\prod_{i=1}^m \tilde{G}_{ins}(\tau_{P^{i-1}(m)}-\tau_{P^i(m)}') \\ \times&\prod_{i=1}^{m-1} \tilde{\mathcal{I}}'_{\nu=0}(\tau'_{P^i(m)}-\tau_{P^i(m)})
    \end{aligned}
     \nonumber \\
    &
    \begin{aligned}
        =&\sgn(\tau'_m-\tau_m) \left[\tilde{G}_{ins}(\tau_{m}-\tau_{P(m)}')
        \right.\\
        &\times\tilde{\mathcal{I}}'_{\nu=0}(\tau'_{P(m)}-\tau_{P(m)})\tilde{G}_{ins}(\tau_{P(m)}-\tau_{P^2(m)}') \cdots\\
        &\times\left.\tilde{\mathcal{I}}'_{\nu=0}(\tau'_{P^{-1}(m)}-\tau_{P^{-1}(m)})\tilde{G}_{ins}(\tau_{P^{-1}(m)}-\tau_{m}') \right],
    \end{aligned}
\end{align}
where we defined $\tilde{\mathcal{I}}'_{\nu=0}(\tau)\equiv\sgn(\tau)\tilde{\mathcal{I}}(\tau)$, and in the last line we have used the fact that $P^{m-1}=P^{-1}$, since $P$ is cyclic.

Our final approximation is to relax the  constrains on $\{\tau'_{\ell_i}\}$ corresponding to time ordering ($\tau'_{\ell_i}<\tau'_{\ell_{i+1}}$) and the non-overlapping instanton condition [see discussion below Eq. \eqref{eq:single_vertex_contr}]. 
This is allowed because $\tilde{G}_{ins}(\tau_{k_i}-\tau_{\ell_i}')$ confines $|\tau_{k_i}-\tau_{\ell_i}'|$ to be at most of order $1/u$. Since the dominant long-time contribution to Eq. \eqref{eq:single_vertex_contr} comes from $\{\tau_{k_i}\}$'s such that $\tau_{k_{i+1}} - \tau_{k_i}$ is large compared to $1/u$, $\tilde{G}_{ins}(\tau_{k_i}-\tau_{\ell_i}')$ ensures that the constrains on $\tau'_{\ell_i}$ are satisfied up to deviations of the order of $\frac{1}{\beta E_\pm}$ -- the small parameter associated with thermal charge fluctuations.

\begin{figure*}
    \centering
    \includegraphics[width=1.\linewidth]{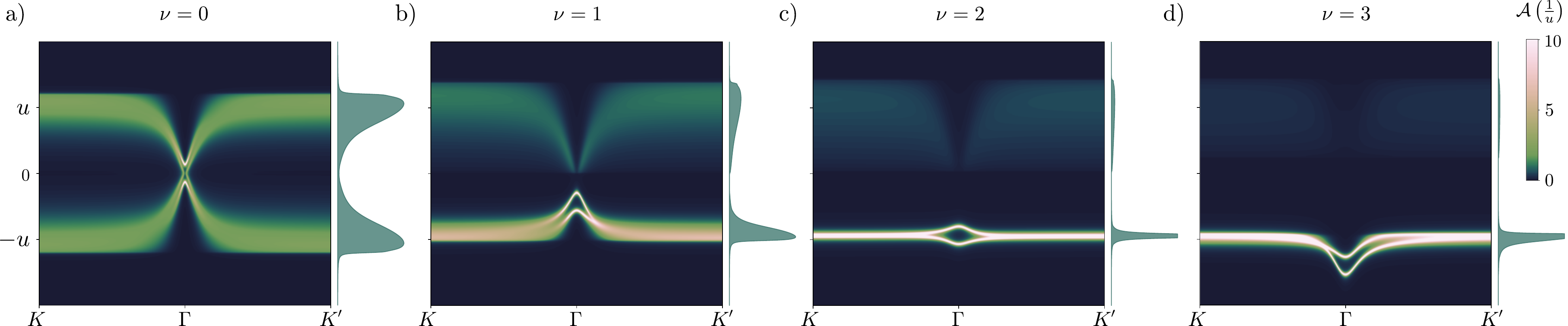}
    \caption{Self-consistently solved spectral function, as defined in Eq.~\eqref{eq:spectral_func}, extracted from Eq.~\eqref{eq:most_div_resummed},\eqref{eq:hybrid_int} and \eqref{eq:D_Gc_relation}, plotted for all integer fillings. To the right of each pannel we plot the integrated spectral function $\mathcal{A}_\text{loc}(\omega)\equiv\int_\k \mathcal{A}(\k,\omega)$, normalized relative to its maximal value.}
    \label{fig:spectral}
\end{figure*}

Summing over all possible cyclic choices of $P$ to account for all non-vanishing creation-annihilation instanton flavor pairings gives rise to a multiplicative $(m-1)!$ factor, canceling the prefactor in Eq.~\eqref{eq:single_vertex_contr}.
Finally, we use \eqref{eq:f_multipoint}, \eqref{eq:convolution_form} and the convolution theorem, to express Eq.~\eqref{eq:single_vertex_contr} as
\begin{equation}
    D_{m-1}(i\omega)=\frac{2}{i\omega}\circledast\left[ G_{ins} \left( \mathcal{I}'_{\nu=0} G_{ins}\right)^{m-1}\right].
\end{equation} 
$G_{ins}$ and $\mathcal{I}'_{\nu=0}$ are the Fourier transforms of $\tilde{G}_{ins}$ and $\tilde{\mathcal{I}}'_{\nu=0}$, respectively, and both are implicitly taken to be at frequency $i\omega$. 
`$\circledast$' denotes convolution in Matsubara-frequencies, where $\frac{2}{i\omega}$ is the Fourier of $\sgn(\tau)$.
Resumming the contribution to all orders in $m$ we find 
\begin{align}
    \label{eq:most_div_nu0}
    D^{(\nu=0)}_\text{div}(i\omega)&=\sum_{m=1}^\infty D_{m-1}(i\omega)\nonumber\\
    &=\left(\frac{2}{i\omega}\right)\circledast\frac{1}{\left[G_{ins}(i\omega)\right]^{-1}-\mathcal{I}'_{\nu=0}(i\omega)}.
\end{align}

In order to generalize this to the case of $\nu\ne0$ we need to account for the asymmetric probability to find a certain flavor empty or filled. This is done by replacing $\sgn(\tau'_i-\tau_i)\rightarrow \sgn(\tau'_i-\tau_i)+\frac{2\nu}{N_f}$ in Eq.~\eqref{eq:f_multipoint},\eqref{eq:convolution_form}. We define $\tilde{\mathcal{I}}'_{\nu}(\tau)=\left(\sgn(\tau)+\frac{2\nu}{N_f}\right)\tilde{\mathcal{I}}(\tau)$ and its Fourier transform $\mathcal{I}'_{\nu}(i\omega)$, to find the expression for $D$ for a generic filling $\nu$
\begin{equation}
    \label{eq:most_div_resummed}
    D_\text{div}(i\omega)=-\left(\frac{2\nu}{N_f}-\frac{2}{i\omega}\circledast\right)\frac{1}{\left[G_{ins}(i\omega)\right]^{-1}-\mathcal{I}'_{\nu}(i\omega)}.
\end{equation}
One can verify, using the definition of $G_{ins}$, that by setting $\mathcal{I}'_{\nu}=0$ we reproduce the tree-level (Hubbard I) expression $D=G_f^{(0)}$.

Together, Eq.~\eqref{eq:most_div_resummed},\eqref{eq:hybrid_int} and \eqref{eq:D_Gc_relation} define a close set of equations for $D_\text{div}(i\omega)$, $\mathcal{I}(i\omega)$, and $G_c(i\omega)$. $D_\text{div}(i\omega)$ depends on the hybridization integral $\mathcal{I}(i\omega)$ through $\mathcal{I}'_{\nu}$. 
In the other direction, $\mathcal{I}(i\omega)$ depends on $D_\text{div}(i\omega)$ through the dependence of $G_c(\k,i\omega)$ on $D(\k,i\omega)$, where we identify $D(\k,i\omega)\equiv D_\text{div}(i\omega)$ to leading order. $D_\text{div}(i\omega)$ must be solved for self-consistently through Eq.~\eqref{eq:most_div_resummed},\eqref{eq:hybrid_int} and \eqref{eq:D_Gc_relation} as opposed to a ``one-shot" calculation where $D_\text{div}(i\omega)$ is calculated using the value of $\mathcal{I}(i\omega)$ within the Hubbard-I approximation. 
The self-consistency amounts to resumming the divergent terms that arise near the Mott bands upon analytical continuation to real frequency in a consistent fashion (see Fig.~\ref{fig:SE_resummation}.c), giving rise to a regular result.  
 
To avoid performing analytical continuation numerically, we analytically continue the self-consistency equations and solve them directly in real frequencies. In App.~\ref{App:analytical_conv} we show that under Wick rotation convolution with $\frac{2}{i\omega}$ transform the imaginary part as follows
\begin{equation}
    \frac{2}{i\omega}\circledast F(i\omega) 
    \xrightarrow[\text{Im. part}]{i\omega\rightarrow\omega+i\eta}
    \sgn(\omega)\operatorname{Im}\left[ F(\omega+i\eta)\right].    
\end{equation}
We use this identity to express the imaginary parts of $\mathcal{I}'_{\nu}(\omega)$ and $D_\text{div}(\omega)$ in real-frequencies
\begin{align}
\label{eq:real_freq_self_consistency}
    \operatorname{Im}\left[ \mathcal{I}'_{\nu}(\omega)\right] &= \left(\sgn(\omega)+\tfrac{2\nu}{N_f}\right) \operatorname{Im}\left[ \mathcal{I}(\omega)\right], \\
    \operatorname{Im}\left[ D_\text{div}(\omega)\right] &=\left(\tfrac{1}{2}-\tfrac{\nu}{N_f}\sgn(\omega)\right)\operatorname{Im}\left[\frac{\sgn(\omega)}{\tfrac{\omega^2-u^2}{2u}-\frac{\mathcal{I}'_{\nu}(\omega)}{2}} \right].\nonumber
\end{align}
The real part is then computed through Kramers-Kronig relations. 

By construction, Eq.~\eqref{eq:convolution_form} does not capture the correct form for instantons that are proximate in time ($|\tau'_i-\tau_i| \ll \frac{1}{u}$). To avoid any artifacts related to short times we employ a soft short-time cutoff on $|\tau'_i-\tau_i|$ by applying a cutoff to $\tilde{\mathcal{I}}(\tau'_i-\tau_i)$ in Eq.~\eqref{eq:convolution_form}. This is implemented as a real-frequency cutoff on Eq.~\eqref{eq:real_freq_self_consistency} 
\begin{equation}
\operatorname{Im}\left[ \mathcal{I}'_{\nu}(\omega)\right]\rightarrow \Theta(\Lambda-|\omega|)\operatorname{Im}\left[ \mathcal{I}'_{\nu}(\omega)\right],
\end{equation}
with $\Lambda>u$.\footnote{This cutoff if motivated by the Lehmann representation, where the imaginary part of the retarded function at frequency $\pm\omega$ translates to the amplitude of the exponentially decaying component $\Theta(\pm\tau)e^{\mp\omega\tau}$ of the imaginary time propagator. Therefore, we omit the terms that contribute mostly at short times.} Given that the cutoff is chosen within the spectral gap ($u<\Lambda\ll\gamma$) the result is independent on the specific choice of cutoff.
In Fig.~\ref{fig:spectral} we plot the spectral function
\begin{equation}
    \label{eq:spectral_func}
    \mathcal{A}(\k,\omega)=-\frac{1}{\pi}\operatorname{Im}\left[ \operatorname{Tr}G_c(\k,\omega) + 
    \operatorname{Tr}G_f(\k,\omega)\right],
\end{equation}
as function of filling, using a self-consistent solution for $D_{\text{div}}(\omega)$. The expressions for $G_c$ and $G_f$ are given in Eq.~\eqref{eq:D_Gc_relation} and~\eqref{eq:D_def} correspondingly.

To provide further insights into our results we consider the chiral limit, away from the $\Gamma$ point, where the spectral function is dominantly of $f$ character. In this case, to leading order in $s^2$ we have $G_f(\k,\omega)\approx D(\omega)$. We can read off the inverse life-time of the quasiparticles from Eq.~\eqref{eq:real_freq_self_consistency}. We find 
\begin{align}
    \label{eq:analytical lifetime}
    \tau^{-1}(\omega)&=-\frac{\sgn(\omega)}{2}\operatorname{Im}\left[ \mathcal{I}'_{\nu}(\omega)\right] \nonumber\\
    &=-\left(\tfrac{1}{2}+\tfrac{\nu}{N_f}\sgn(\omega)\right) \operatorname{Im}\left[ \mathcal{I}(\omega)\right].
\end{align}
We note that Eq.~\eqref{eq:analytical lifetime} has a factor of $\tfrac{1}{2}+\tfrac{\nu}{N_f}$ for `doublon' ($\omega>0$) excitations and $\tfrac{1}{2}-\tfrac{\nu}{N_f}$ for `holon' ($\omega<0$) excitations. This is exactly the phase space for an excitation of the opposite type, meaning $\tau^{-1}_\text{doublon}\propto Z_\text{holon}$ and vice versa.
This result has a simple physical interpretation. 
An excitation composed of an added $f$ electron can decay by creating a particle-hole pair, where the particle is a $c$ electron and the hole is an $f$ electron of a different flavor on the same site. To conserve energy, the energy of the $c$ electron must be of the order of the Mott band energy. 
The phase space of this process is proportional to the number of excitable hole states (i.e., the number of occupied $f$ electrons) on the site. 
Conversely, an initial $f$ hole excitation decays by creating an $f$ particle and a $c$ hole. 
This leads to a particle-hole asymmetry in the width of the Mott bands when $\nu\ne 0$, as is visible in Fig.~\ref{fig:spectral}. We note that $\operatorname{Im}\mathcal{I}(\omega)$ itself is particle-hole asymmetric when $\nu \ne 0$; however, we find that the asymmetry in $\operatorname{Im}\mathcal{I}(\omega)$ has the same sign as that of the prefactor of Eq. \eqref{eq:analytical lifetime}. This asymmetry between the width of the two Mott bands is seen also in DMFT~\cite{rai2024dynamical,datta2023heavy}.

Finally, specializing to half-filling ($\nu=0$), we analytically continue $\mathcal{I}$ from Eq. \eqref{eq:I} to real frequencies, and obtain the simple expression
\begin{equation}
    \tau^{-1}(\pm\omega_\text{Mott})=\frac{\pi}{2}s^2N_fu+\mathcal{O}\left(s^4u\right),
\end{equation}
where $\omega_\text{Mott}=u\sqrt{1-s^2}$ is defined as the band-edge within the Hubbard I approximation.

\subsection{Excluded Diagrams}
\label{section:excluded}

We now consider the non-local diagrams excluded from our resummation scheme, and argue that these are indeed small compared to the terms included in the resummation.  
We focus on the excluded diagram with the smallest number of fermionic loops, shown in Fig.~\ref{fig:SE_resummation}c. 
As we shall demonstrate, this diagram is indeed suppressed compared to the term kept in the resummation with the same number of loops, as long as the temperature is not too low. At asymptotically low temperature, our approximation fails due to a $\tfrac{1}{T}$ divergence of this diagram associated with a diverging flavor correlation length.

Using the approximate form of the $\Gamma^{(2)}$ vertex found in Sec.~\ref{section:4f}, we find the contribution to $\left[D_{\lambda}(\k,i\omega)\right]_{b_1,b_2}$ from this diagram to be (see App.~\ref{App:excluded} for derivation)
\begin{align}
&\left[\left(\tfrac{2\nu}{N_f}\right)^2-1\right]\left(\frac{u}{u^2-(i\omega+\delta\mu)^2}\right)^2 \frac{1}{T} \sum_{\lambda'}\int_{\k'}
\nonumber\\ 
&\times\left[\hat{\Theta}_{\k-\k'}^{(\lambda,\lambda')}\right]_{(b_1,b_3;b_2,b_4)}
\left[\hat{\gamma}_{\k'}G_{c,\lambda'}\left(\k',i\omega\right)\hat{\gamma}_{\k'}^\dagger\right]_{b_3,b_4},
\label{eq:excluded_full}
\end{align}
where the particle-hole bubble $\hat{\Theta}_{\k-\k'}^{(\lambda,\lambda')}$ is given by
\begin{align}
\label{eq:Theta_c}
    &\left[\hat{\Theta}_{\q}^{(\lambda,\lambda')}\right]_{(b_1,b_3;b_2,b_4)} \equiv\left[\left(\tfrac{2\nu}{N_f}\right)^2-1\right]\int_{\omega}\left(\tfrac{u}{u^2-(i\omega+\delta\mu)^2}\right)^2
     \nonumber\\
    &\begin{aligned}
        \times\int_{\k'}&\left[\hat{\gamma}_{\k'+\q}G_{c,\lambda}\left(\k'+\q,i\omega\right)\hat{\gamma}_{\k'+\q}^\dagger\right]_{b_1,b_2}\\
        &\times\left[\hat{\gamma}_{\k'}G_{c,\lambda'}\left(\k' ,i\omega\right)\hat{\gamma}_{\k'}^\dagger\right]_{b_4,b_3}.
    \end{aligned}
\end{align} 
The notation of $\hat{\Theta}_\q^{(\lambda,\lambda')}$ is chosen to indicate the relation of this quantity to the Curie temperature of the system as we will show in the next section. Using a bound on the norm\footnote{We take the norm on $\hat{\Theta}_\q^{(\lambda,\lambda')}$ to be the operator norm given by its maximal eigenvalue.} $\Theta_c = \max_\q ||\hat{\Theta}_\q^{(\lambda,\lambda')}||$, given by the actual Curie temperature of the model, we can bound the contribution to $D(i\omega)$ from this diagram by
\begin{equation}
    \frac{\Theta_c}{T}\left[1-\left(\frac{2\nu}{N_f}\right)^2\right]\left(\frac{u}{u^2-(i\omega+\delta\mu)^2}\right)^2 \mathcal{I}(i\omega),
    \label{eq:excluded}
\end{equation} 
which is suppressed at high temperatures relative to the one-loop  contribution (Eq.~\ref{eq:D1}) by the factor of $\tfrac{\Theta_c}{T}$. In the flat-chiral limit we find $\Theta_c\sim s^2u$ (see Sec.~\ref{section:susceptibility}), therefore our results are applicable over a wide range of temperatures, $\Theta_c \ll  T \ll u$. For realistic parameters for MATBG, we find $\Theta_c$ to have a similar magnitude away from the chiral limit (App.~\ref{App:susceptibility}), ensuring a scale separation between $\Theta_c$ and $u$ both in and away of the chiral limit.

As we claimed previously, this non-local diagram has a weaker real-frequency divergence in $\omega\mp E_\pm$ compared to the other two-loops diagrams found in Fig.~\ref{fig:SE_resummation}c. Therefore, its omission from the resummation for the Mott band self-energy is justified.

\section{Flavor Susceptibility}
\label{section:susceptibility}

\begin{figure}
    \centering
    \includegraphics[width=1.\linewidth]{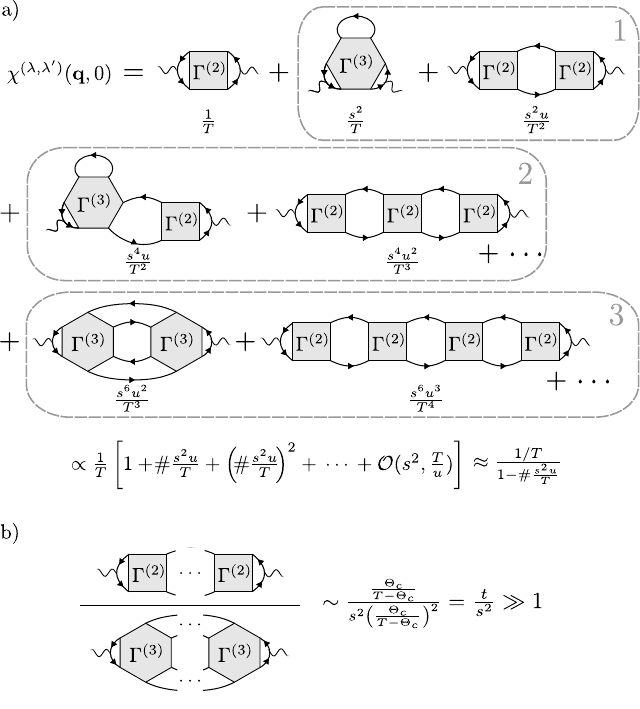}
    \caption{a) Example of diagrams contributing to the flavor susceptibility to order one, two and three loops. Below each diagram we specify its scaling with $s^2$, given by the number of loops, and with $\frac{u}{T}$, given by the number of unconstrained imaginary time integrations.
    After we identify the existence of $\left(\frac{u}{T}\right)^{n_{\text{loops}}}$ divergent diagrams, we assume a scaling of $\frac{u}{T}\sim s^2$ to reorganize the expansion in the bottom line of this panel. This results in a Curie-Weiss form. All of the terms of order one in this resummation are of the ladder type (the rightmost diagrams in each line). b) By comparing the ladder resummation from panel (a) to a different infinite set of diagrams, we estimate the reduced temperature $t_{\rm{MF}}=\frac{T-\Theta_c}{\Theta_c}$ at which corrections to the mean-field (Curie-Weiss) form of the susceptibility become significant. This only occurs for parametrically small $t_{\rm{MF}}\sim s^2$. One can see that this is the first correction that become significant at small $t_{\rm{MF}}$ as it is composed of two independent ladder resummations and therefore a higher power of $t_{\rm{MF}}^{-1}$.}
    \label{fig:susceptibility_resummation}
\end{figure}

We define the flavor susceptibility as 
\begin{equation}
    \chi^{(\lambda,\lambda')}_{\a',\a}(q)=\left\langle \left(S_{\a'}^{(\lambda,\lambda')}(q)\right)^\dagger S_\a^{(\lambda,\lambda')}(q)\right\rangle,
\end{equation}
with $S_\mathbf{a}^{(\lambda,\lambda')}(q)$ the flavor-moment density at momentum and frequency $q=(\q,\Omega)$ and particle-hole orbital indices given by $\a=(a_1,a_2)$, where $a_1,a_2$ can be orbital indices of either $f$ or $c$ electrons ($a_i=1,\dots,6$). To capture the temperature dependence, it is sufficient to consider only flavor moments of $f$ orbital character ($b_i=1,2$), for which $S_\b^{(\lambda,\lambda')}(q)=\int \frac{d\tau}{\sqrt{\beta}} e^{i\Omega\tau}\sum_\R\frac{e^{i\q\cdot\R}}{N_{s}}\bar{f}_{\R,\lambda,b_1}(\tau)f_{\R,\lambda',b_2}(\tau)$. 

To zero loops order the flavor susceptibility is given by the single-site susceptibility, having a Curie form
\begin{align}
    &\chi^{(\lambda,\lambda')}_{\b',\b}(q)\nonumber\\
    &= \int_{\tau,\tau'} \frac{e^{i\Omega(\tau-\tau')}}{\beta}\braket{\bar{f}_{\lambda',b'_2}(\tau')f_{\lambda,b'_1}(\tau')\bar{f}_{\lambda,b_1}(\tau)f_{\lambda',b_2}(\tau)}_{0}\nonumber\\
    &=\int_{\delta\tau} e^{i\Omega\delta\tau} \frac{1-\left(\tfrac{2\nu}{N_f}\right)^2}{4(1-N_f^{-1})}\delta_{\b,\b'}\nonumber\\
    &=\frac{1-\left(\tfrac{2\nu}{N_f}\right)^2}{4(1-N_f^{-1})}\frac{\delta_{\Omega,0}}{T}\delta_{\b,\b'},
\end{align}
with $\delta_{\b,\b'}\equiv\delta_{b_1,b'_1}\delta_{b_2,b'_2}$. In the third line we took the limit of temperatures far below the charging energy $\beta E_\pm\gg1$.

\subsubsection{One-Loop Corrections}
Within one-loop order we have two contributing diagrams, given in the first line of Fig.~\ref{fig:susceptibility_resummation}.a. We can infer the maximal power of $\frac{1}{T}$ contributed by each diagram by considering the number of free (unconstrained) times we integrate over. For example: the zero-loops contribution has a term independent of $\delta\tau=\tau-\tau'$, and therefore scales as $\frac{1}{T}$. The first one-loop diagram, composed of a single $\Gamma^{(3)}$ vertex, has no additional powers of $\frac{1}{T}$, as the two additional internal times are constrained to be close to each other and to one of the external times. This diagram amounts to a renormalization of the local moment. The same counting tells us that the second one-loop diagram is of order $\frac{1}{T^2}$, since all of the four additional internal times are constrained to be close to each other, but are unconstrained relative to the two external times. This diagram's contribution is given by
\begin{equation}
    \frac{1-\left(\tfrac{2\nu}{N_f}\right)^2}{4(1-N_f^{-1})}\frac{\delta_{\Omega,0}}{T}\frac{\left[\hat{\Theta}_{\q}^{(\lambda,\lambda')}\right]_{(\b;\b')}}{T}\left( 1-\mathcal{O}(\tfrac{T}{u})\right),
    \label{eq:one_loop_susceptibility}
\end{equation}
where $\hat{\Theta}_{\q}^{(\lambda,\lambda')}$ is the Curie temperature matrix defined in Eq.~\eqref{eq:Theta_c}. In the flat-chiral limit $\hat{\Theta}_{\q}^{(\lambda,\lambda')}$ is independent of $\lambda,\lambda'$ and is proportional to an identity matrix $\left[\hat{\Theta}_{\q}^{(\lambda,\lambda')}\right]_{(\b;\b')}=\delta_{\b,\b'}\Theta_\q^{\text{(flat-chiral)}}$. At charge neutrality it is given by
\begin{align}
\label{eq:T_Curie_flat-chiral}
    \begin{aligned}
    \Theta_\q^{\text{(flat-chiral)}}=&\int_{\omega,k}\left(\tfrac{u}{u^{2}+\omega^{2}}\right)^{2}\operatorname{Tr}\left[\gamma^2\tfrac{1+\zeta_{z}}{2}G_{c}\left(\mathbf{k}+\tfrac{\mathbf{q}}{2},i\omega\right)\right]\\
    &\times\operatorname{Tr}\left[\gamma^2\tfrac{1+\zeta_{z}}{2}G_{c}\left(\mathbf{k}-\tfrac{\mathbf{q}}{2},-i\omega\right)\right]
    \end{aligned}\nonumber\\
    \begin{aligned}
    =&\int_{\omega,k}\frac{\gamma^{2}u}{i\omega\left(\gamma^{2}+u^{2}+\omega^{2}\right)-\frac{u^{2}+\omega^{2}}{i\omega}v_{\star}^{2}\left|\mathbf{k}+\frac{\mathbf{q}}{2}\right|^{2}}\\
    &\times \frac{\gamma^{2}u}{-i\omega\left(\gamma^{2}+u^{2}+\omega^{2}\right)-\frac{u^{2}+\omega^{2}}{-i\omega}v_{\star}^{2}\left|\mathbf{k}-\frac{\mathbf{q}}{2}\right|^{2}}.
    \end{aligned}
\end{align}
It is maximal for $\q=0$, where $\Theta_c=\Theta_{\q=0}$ is given by
\begin{align}
\label{eq:Curie_flat_chiral}
    \Theta_{\q=0}^{\text{(flat-chiral)}}&=\int_{\omega,k}\left(\frac{\gamma^{2}u}{\omega}\right)^{2}\frac{1}{\left[\gamma^{2}+u^{2}+\omega^{2}+\frac{u^{2}+\omega^{2}}{\omega^{2}}v_{\star}^{2}\left|\mathbf{k}\right|^{2}\right]^{2}}
    \nonumber\\
    &\approx\frac{1}{2}s^{2}u\left(1-\frac{u}{\sqrt{u^{2}+\gamma^{2}}}\right),
\end{align}
where we neglected corrections of order $s^4$ by taking the momentum cutoff in the integral to infinity. 
We therefore established that in the flat-chiral and projected limit we have $\Theta_c\sim s^2u$, whereas in the opposite limit of $\gamma\ll u$ we find $\Theta_c\sim s^2\frac{\gamma^2}{u}$. This correspond to the expected RKKY coupling in this limit $J_{RKKY}\sim J_K^2\nu_c(\omega\sim u)$, where $J_K=\frac{\gamma^2}{u}$ and $\nu_c(\omega)\sim\frac{s^2|\omega|}{\gamma^2}$ is the bare (Dirac) $c$-electron density of states.

Thus, at one-loop order we found a diagram that diverges with temperature as $1/T^2$, which is faster than the $1/T$ divergence of the zero-loops expression. As a consequence the one-loop and zero-loop contributions become comparable at temperatures $T\sim\Theta_c\sim s^2u$. At temperatures of this order we have no reason to stop at finite-loop order. We must consider all the diagrams that give a comparable contribution at any loop-order to calculate the susceptibility in a consistent manner.

\subsubsection{Ladder-Resummation}
We assume that $T\sim s^2u$, and search for all diagrams which contribution scale as $\frac{1}{T}\left(\frac{s^2u}{T}\right)^{n_\text{loops}}\sim \frac{1}{T}$. As mentioned before, the power of $1/T$ can be identified by counting the number of unconstrained times integrated over, whereas the power of $s^2$ is the number of loops. In Fig.~\ref{fig:susceptibility_resummation}.a we specify the scaling of each of the new type of diagrams appearing at order one, two and three loops. We claim that to any loop-order, the only diagram scaling as $\frac{1}{T}\left(\frac{s^2u}{T}\right)^{n_\text{loops}}$ is of the ladder type. A resummation of these diagrams can be done exactly up to order $\tfrac{1}{\beta E_\pm}$ corrections. It results in a Curie-Weiss susceptibility
\begin{align}
\label{eq:Curie-Weiss}
    \hat{\chi}_{\mathbf{b},\mathbf{b}'}^{(\lambda,\lambda')}(q)&= \left[\frac{\chi_\text{single-site}}{1-\beta\hat{\Theta}_{\q}^{(\lambda,\lambda')}}\right]_{\mathbf{b},\mathbf{b}'}
    \nonumber\\
    &=\frac{1-\left(\tfrac{2\nu}{N_f}\right)^2}{4(1-N_f^{-1})}\left[\frac{\delta_{\Omega,0}}{T-\hat{\Theta}_{\q}^{(\lambda,\lambda')}}\right]_{\mathbf{b},\mathbf{b}'},
\end{align}
with $\hat{\chi}^{(\lambda,\lambda')}(q)$ and $\hat{\Theta}_{\q}^{(\lambda,\lambda')}$ matrices with indices of $f$ particle-hole orbitals $\b$.

The eigenvalues of $\hat{\Theta}_{\q}^{(\lambda,\lambda')}$ correspond to the Curie temperature attributed to different flavor orders.\footnote{At charge-neutrality, in the presence of a finite mass ($M\ne0$), $\hat{\Theta}_{\q}^{(\lambda,\lambda')}$ itself is temperature dependent. It has a logarithmic divergence with temperature due to the quadratic band-touching. This results in a logarithmic correction to the Curie-Weiss form of the susceptibility.} In App.~\ref{App:susceptibility} we plot the eigenvalues of $\hat{\Theta}_{\q}^{(\lambda,\lambda')}$ for different limits of the model. Our results reproduce the hierarchy of lifted degeneracies previously predicted for the zero-temperature flavor orders in~\cite{bultinck2020ground}. As in the zero-temperature case, accidental degeneracies between different valley-diagonal and valley-off-diagonal ordering tendencies arise in the flat and/or chiral limits with a maximal degeneracy when both limits are taken together. We note that in the flat-chiral limit we find the susceptibility to be $SU(8)$ symmetric, which is even more symmetric than the expected $SU(4)\times SU(4)$. We attribute this to the fact that we only consider Hubbard interaction while omitting other symmetry allowed interactions such as a direct $f-c$ exchange interaction.

We emphasize the fact that for a generic heavy fermion model (relevant, e.g., to heavy-fermion compounds), the ladder resummation is not justified. It is only due to the existence of the small parameter $s$ that this resummation is controlled down to low temperatures (of the order of the Curie temperature). In the remainder we discuss the breakdown of this resummation upon further lowering of the temperature.

\subsubsection{Critical fluctuation Regime}
Universality of critical behavior tells us that Eq.~\eqref{eq:Curie-Weiss} cannot be correct all the way down to the phase transition at $t\equiv\frac{T-T_c}{T_c}=0$.
Motivated by this fact we look for sets of diagrams which amount to non-negligible corrections to the susceptibility at small reduced temperatures. To find these diagrams we replace bubbles with ladder resummations which has the effect of taking $\frac{\Theta_c}{T}\rightarrow\frac{\Theta_c}{T-\Theta_c}\equiv t_{\rm{MF}}^{-1}$ (Fig.~\ref{fig:susceptibility_resummation}.b). We therefore find that diagrams with multiple independent ladder resummations diverges faster with $t_{\rm{MF}}^{-1}$ and become order one when $t_{\rm{MF}}\sim s^2$. 

Another contribution which we neglect and becomes significant as $t_{\rm{MF}}\rightarrow0$ is the dressing of the fermionic quasiparticles by flavor-fluctuations. To leading order, this kind of dressing is captured by the rightmost diagram in Fig.~\ref{fig:SE_resummation}.c. A similar replacement of the single bubble with a ladder re-summation results in a set of diagrams contributing an $s^2t_{\rm{MF}}^{-1}$ relative correction to the $c$-electrons self energy at low energies, thus affecting the strongly fluctuating regime in the presence of low energy $c$-like excitations. Still, the effect is significant only for $t_{\rm{MF}}\lesssim s^2$.

We conclude that while critical behavior is unavoidable we can bound the critical regime to a parametrically small reduced temperature range $t_{\rm{MF}}\sim s^2$ above which Eq.~\eqref{eq:Curie-Weiss} is a good approximation. Furthermore, unlike the local moment fluctuations, these finite-range flavor fluctuations are not $SU(N_f)$ symmetric, hence the small parameter to bound the extent of the critical regime is indeed $s^2$ and not $s^2N_f$ as in the rest of the expansion. In the case of an accidental degeneracy between multiple flavor orders, as in the case of the flat and/or chiral limit, the appropriate small parameter will be $s^2N_d$, with $N_d$ the number of degenerate orders corresponding to the maximal Curie temperature.

\section{Discussion \& Outlook}
\label{section:discussion}

In this work we developed a strong-coupling diagrammatic approach to heavy-fermion models controlled in the chiral limit by a small phase space parameter $s^2N_f$, where we identified the number of $c$ electron loops with powers of the small parameter. Away from the chiral limit the expansion is controlled by two small energy scales: The hybridization function $\mathcal{I}\ll U$, and the Curie temperature $\Theta_c\ll T$.

We showed that at sufficiently high temperatures, the self-energy in such models is approximately local in space. To zeroth order in the small parameter we showed that the self-energy is identical to that of the Hubbard-I approximation, and to the self-energy found in~\cite{ledwith2024nonlocal} in the chiral and projected limit of the model.
By taking the limit of large $N_f$ while keeping $\mathcal{I}$ constant, we derived the quasi-particle lifetime both for the dispersive part of the spectrum ($|\omega|\ll U$) and for the flat, Mott-like, part ($|\omega| \sim U/2$), finding an appreciable inverse lifetime of the order of $s^2N_f$. 

As the temperature is lowered, $T\rightarrow\Theta_c$. 
We found that non-local contributions to the self-energy become appreciable, with the leading non-local contribution proportional to $\frac{\Theta_c}{T}$. 
We further studied the flavor-susceptibility of the model at intermediate temperatures $U\gg T\gtrsim \Theta_c$, to find that it obeys a Curie-Weiss behavior up to corrections of order $s^2\tfrac{T-\Theta_c}{\Theta_c}$. We thus infer that the crossover from a Curie-Weiss behavior to a strongly fluctuating critical regime occurs only 
when $\tfrac{T-\Theta_c}{\Theta_c}$ is of the order of $s^2$. 
This property originates from an emergent mean-field behavior in the model, associated with the fact that the effective interactions between $f$-moments are long ranged in space (with a characteristic range of $\sim 1/s$ lattice spacings).
This property allows for a controlled analysis of the problem down to low temperatures, parametrically close to the critical regime.
 
The emergent spatial locality of the self-energy in the limit $\mathcal{I}\ll U$ and $\Theta_c\ll T$ implies that DMFT is asymptotically exact in this regime. This allows in principle numerical computation of correlators in imaginary-time, beyond the large $N_f$ limit assumed in our derivation. Moreover, the diagrams we identify as the dominant ones are only a subset of all the local diagrams. A specialized algorithm to efficiently compute only those dominant diagrams in the small $s$ limit can be developed based on the hybridization expansion solver~\cite{werner2006continuous,werner2006hybridization}.

An important but straightforward extension of our results is to include additional interactions existing in the topological heavy fermion model,  specifically the $f-c$ direct and exchange interactions.
Another, more intriguing extension would be to consider the temperature dependent corrections to the single-body self energy above the Curie temperature. The fact that the susceptibility has a simple Curie-Weiss form suggests that this can be carried out analytically, and may shed light on the temperature dependence of the resistivity in MATBG~\cite{cao2020strange,jaoui2022quantum}.
\vspace{1em}
\paragraph*{Note added} -- We became aware of several upcoming works on closely related topics~\cite{Pavel_2026,hu_to_appear_2026,Nemin2026}. Our results agree with theirs where they overlap.

\begin{acknowledgments}
We thank Shahal Ilani, Eyal Keshet and Jiewen Xiao for collaboration on related projects. We are also grateful to Elena Bascones, B. Andrei Bernevig, Antoine Georges, Leonid Glazman, Andrew Hardy, Haoyu Hu, Shahal Ilani, Eslam Khalaf, Patrick Ledwith, Pavel Nosov, Yuval Oreg, Giorgio Sangiovanni, André-Marie Tremblay, Oskar Vafek, and Roser Valent\'{\i} for useful discussions. We are especially grateful to Eyal Keshet for valuable discussions and for collaboration on closely related topics. 

This work was supported by NSF-BSF Award No. DMR-2310312, the Simons Foundation Collaboration on New Frontiers in Superconductivity (Grant SFI-MPS-NFS-00006741-03), and CRC 183 of the Deutsche Forschungsgemeinschaft (Project C02). 
We thank the hospitality of the Kavli Institute for Theoretical Physics (KITP)
supported by grant NSF PHY-2309135 and the Gordon and Betty Moore Foundation Grant No. 2919.02, where this work was conceived. 


\end{acknowledgments}

\appendix
\begin{widetext}
\section{Topological Heavy Fermion Model}
\label{App:THF_model}

The complete THF model~\cite{song2022magic} is given by
\begin{equation}
    H = H_c+H_{fc}+H_U+H_J+H_V+H_W,
\end{equation}
with
\begin{align}
    H_c &=\sum_{\k,\lambda}c^\dagger_{\k,\lambda,a}\left[ \hat{h}_\k^{(c,\lambda)}\right]_{a,a'}c_{\k,\lambda,a'} \\
    H_{fc} &=\sum_{\k,\lambda}f^\dagger_{\k,\lambda,b}\left[\hat{\gamma}_\k\right]_{b,a}c_{\k,\lambda,a} +\text{h.c.} \\
    H_U &= \frac{U}{2}\sum_\R \delta n_\R^2 \\
    H_J &= -J\sum_{\R\q}\sum_{\mu\nu}\sum_{\xi=\pm}e^{-i\q\cdot\R}:\hat{\Sigma}^{(f,\xi)}_{\mu\nu}(\R)::\hat{\Sigma}^{(c,\xi)}_{\mu\nu}(\q): \\
    H_W &=W\sum_{\R\q}e^{-i\q\cdot\R}\delta n_\R\left(\frac{\sum_{\k,a,\lambda}}{N_s}:c^\dagger_{\k+\q,\lambda,a}c_{\k,\lambda,a}:\right)\\
    H_V &=\frac{1}{2N_s\Omega_0}\sum_{\substack{\k_1,a_1,\lambda_1\\\k_2,a_2,\lambda_2\\ \q}}V(\q):c^\dagger_{\k_1,\lambda_1,a_1}c_{\k_1+\q,\lambda_1,a_1}::c^\dagger_{\k_2+\q,\lambda_2,a_2}c_{\k_2,\lambda_2,a_2}:,
\end{align}
where $N_s$ is the number of moir\'{e} unit cells and $\Omega_0$ is the unit cell area. The one-body Hamiltonian terms $\hat{h}_\k^{(c,\lambda)}$ and $\hat{\gamma}_\k$ are defined in the main text (Eq.~\ref{eq:H_c},\ref{eq:gamma_def}).
The flavor moments appearing in $H_J$ are defined as 
\begin{align}
    \hat{\Sigma}^{(f,\xi)}_{\mu\nu}(\R)&=\left[A^{(\xi)}_{\mu\nu}\right]_{b,\lambda;b',\lambda'}f^\dagger_{\R,\lambda,b}f_{\R,\lambda',b'}, \\
    \hat{\Sigma}^{(c,\xi)}_{\mu\nu}(\q)&=\left[B^{(\xi)}_{\mu\nu}\right]_{a,\lambda;a',\lambda'}\frac{\sum_\k}{N_s}c^\dagger_{\k+\q,\lambda,a}c_{\k,\lambda',a'},
\end{align}
where 
\begin{align}
    A^{(\xi)}_{\mu\nu}&=\frac{\xi+\sigma_z\tau_z}{4}\times\left\{\sigma_0\tau_0,\sigma_y\tau_x,\sigma_y\tau_y,\sigma_0\tau_z\right\}\times s_\nu, \\
    B^{(\xi)}_{\mu\nu}&=\frac{\zeta_0-\zeta_z}{2}\frac{\xi+\sigma_z\tau_z}{4}\times\left\{\sigma_0\tau_0,-\sigma_y\tau_x,-\sigma_y\tau_y,\sigma_0\tau_z\right\}\times s_\nu,
\end{align}
with $s_\mu$ acting on spin, $\tau_\mu$ acting on valley and $\sigma_\mu,\zeta_\mu$ acting on orbital space such that $\sigma_\mu$ act both on $f$ and $c$ electrons, and $\zeta_\mu$ act on an orbital degree of freedom which is unique to the $c$ electrons (see also discussion around Eq.~\ref{eq:H_c}-\ref{eq:reduced_Hc}).

Within this work we take $J=0$ and treat $H_W,H_V$ within the Hartree approximation. The parameters used to produce the figures in the text are given in Tab.~\ref{tab:H_params}, where $\gamma$ and $U$ were chosen to match the band gap and Mott resonance observed in experiment~\cite{xiao2025interactingenergybandsmagic}, respectively. $v_\star$ and $v_\star'$ were scaled compared to the values of Ref.~\cite{song2022magic} in order to keep $N_fs^2$ sufficiently smaller than one. $W$ and $V$ were rescaled by a similar factor to $U$ relative to Ref.~\cite{song2022magic}.

\begin{table}[H]
    \centering
    \begin{tabular}{c|c|c|c|c|c|c}
        $v_\star$ & $M$ & $\gamma$ & $v_\star'$ & $U$ & $W$ & $\frac{1}{\Omega_0}V(\q=0)$ \\
         $-1.6$eV$\cdot$nm & $2$meV & $-60$meV & $250$meV$\cdot$nm & $30$meV & $23$meV & $23$meV
    \end{tabular}
    \caption{Hamiltonian parameters}
    \label{tab:H_params}
\end{table}

\section{Loop Power Counting in the Flat-Chiral Limit}
\label{App:loop_counting}
We show that in the projected, flat-chiral limit, any diagram contributing to the
$c$-electron self-energy $\Sigma_c$ at $n_l$-loop order satisfies
\begin{equation}
    \Sigma_c^{(n_l)} \;\sim\; \frac{\gamma^2}{u} F\left(\frac{T}{u}\right) \,s^{2n_l},
    \label{eq:loop_scaling}
\end{equation}
up to logarithmic corrections in $s$ and combinatorial prefactors, where $F(T/u)$ is a dimensionless scaling function. Away from the projected limit there is an additional smooth dependence on $u/\gamma$.

\subsection{Setup: general self-energy diagram}
A general diagram contributing to $\Sigma_c$ is built from $n$ interaction
vertices $\Gamma^{(m_1)},\ldots,\Gamma^{(m_n)}$ drawn from the effective
action~\eqref{eq:eff_action}, with $m_i\geq1$.  Each vertex $\Gamma^{(m_i)}$
carries $m_i$ incoming and $m_i$ outgoing $c$-electron legs.  For a
self-energy diagram, two $c$-legs (one incoming, one outgoing) are left
external and all remaining legs are contracted into internal propagators.

\subsection{Counting propagators and loops}
The $n$ vertices contribute a total of $2\sum_{i=1}^n m_i$ $c$-leg
endpoints.  Subtracting the 2 external endpoints and pairing the remainder
into internal $c$-propagators gives
\begin{equation}
    P = \sum_{i=1}^n m_i - 1
    \label{eq:P_count}
\end{equation}
internal $c$-electron propagators.  For a connected diagram the number of
independent loops is then
\begin{equation}
    n_l = P - n + 1 = \sum_{i=1}^n m_i - n.
    \label{eq:nl_count}
\end{equation}

\subsection{Scaling form of $G_c$ in the flat-chiral limit}
In the flat-chiral limit the Hubbard-I $c$-propagator takes the form
(Eq.~\eqref{eq:Gc_H1})
\begin{equation}
    G_{c,\lambda}(\k,i\omega) =
    \frac{-i\omega}{\omega^2\!\left(1+\dfrac{\gamma^2}{\omega^2+u^2}\right)+v_\star^2|\k|^2}.
    \label{eq:Gc_flat_chiral}
\end{equation}
Introducing the dimensionless variables $\tilde\omega=\omega/u$ and
$\tilde\k=\k/k_{BZ}$, and using $s^2=\gamma^2/(v_\star k_{BZ})^2$ where $\pi k_{BZ}^2=A_{BZ}$
(Eq.~\eqref{eq:s2}), the denominator can be rewritten to read
\begin{equation}
    G_{c,\lambda}(\k,i\omega)
    = \frac{u}{\gamma^2}\,
      g_c\!\left(\frac{\tilde\k}{s},\,\tilde\omega,\,\frac{u}{\gamma}\right),
    \label{eq:Gc_scaling}
\end{equation}
where
\begin{equation}
    g_c\!\left(\frac{\tilde\k}{s},\,\tilde\omega,\,\frac{u}{\gamma}\right)
    \equiv
    \frac{-i\tilde\omega}{\,\dfrac{\tilde\omega^2}{(\gamma/u)^2}
    +\dfrac{\tilde\omega^2}{\tilde\omega^2+1}
    +\left|\tilde\k/s\right|^2\,}
    \label{eq:gc_def}
\end{equation}
is a dimensionless function of order unity when its arguments are of order
unity.  In the projected limit $u/\gamma\to0$ at fixed $s$, the first term
in the denominator is suppressed and $g_c$ simplifies to
\begin{equation}
    g_c\!\left(\frac{\tilde\k}{s},\,\tilde\omega,\,0\right)
    =
    \frac{-i\tilde\omega}{\,\dfrac{\tilde\omega^2}{\tilde\omega^2+1}
    +\left|\tilde\k/s\right|^2\,}.
    \label{eq:gc_projected}
\end{equation}
The factor $u/\gamma^2$ in Eq.~\eqref{eq:Gc_scaling} sets the overall
magnitude of $G_c$.

\subsection{Power counting}
We now assemble the scaling of a general self-energy diagram.

\paragraph{Vertices.}
From the definition Eq.~\eqref{eq:Gamma_def}, each vertex $\Gamma^{(m_i)}$
contains $2m_i$ hybridization factors
(contributing $\gamma^{2m_i}$) multiplied by the connected $f$-electron
$2m_i$-point correlator, with units of $(\rm{energy})^{-m_i}$.
Since the only energy scales in the $2m_i$-point correlator are $u$ and $T$, we can write the contribution of the vertex to the diagram as
\begin{equation}
    \Gamma^{(m_i)} \;\sim\; \frac{\gamma^{2m_i}}{u^{m_i}} f_i\left(\frac{T}{u}\right).
    \label{eq:Gamma_scaling}
\end{equation}
The dimensionless function $f_i(T/u)$ depends on how the vertex is embedded in a particular diagram (see discussion below Eq.~\ref{eq:loop_scaling_final}).
\paragraph{Internal propagators.}
Each of the $P$ internal $c$-propagators contributes $G_c\sim u/\gamma^2$
from Eq.~\eqref{eq:Gc_scaling}, with the function $g_c=O(1)$.

\paragraph{Loop integrals.}
From Eq.~\eqref{eq:gc_projected}, $g_c(\tilde\k/s,\tilde\omega)$ is of
order unity only when $|\tilde\k|\lesssim s$, i.e.\ in a fraction $\sim s^2$
of the Brillouin zone.  Rescaling the loop momentum $\q=\tilde\k/s$,
each momentum integral evaluates to
\begin{equation}
    \int\frac{d^2\tilde\k}{A_{BZ}}\,
    g_c\!\left(\frac{\tilde\k}{s},\tilde\omega\right)
    = s^2\int\frac{d^2\q}{A_{BZ}}\,
    g_c\!\left(\q,\tilde\omega\right)
    \;\sim\; s^2,
    \label{eq:loop_integral}
\end{equation}
where the remaining integral over $\q$ is $O(1)$, up to logarithmic factors in $s$.\footnote{Such a logarithmic factor appears in the one-loop diagram; see Eq.~\eqref{eq:I}. Dimensional analysis shows that the divergence of the momentum integrals is never stronger than logarithmic in $s^{-1}$, since the number of internal $c$-propagators (each scaling as $k^{-2}$) is larger or equal to the number of loops.}  The suppression by $s^2$
reflects precisely the small fraction of the Brillouin zone with appreciable
$c$-electron character. 
Each of the $n_l$ loops
therefore contributes one factor of $s^2$. 

\paragraph{Result.}
Collecting all factors:
\begin{align}
    \Sigma_c
    &\;\sim\;
    \underbrace{\prod_{i=1}^n \frac{\gamma^{2m_i}}{u^{m_i}}\,
      f_i\!\left(\frac{T}{u}\right)}_{\text{vertices}}
    \times
    \underbrace{\left(\frac{u}{\gamma^2}\right)^{P}}_{\text{propagators}}
    \times
    \underbrace{\left(s^2\right)^{n_l}}_{\text{loops}}.
    \label{eq:power_count}
\end{align}
The $\gamma$ powers combine as
$\gamma^{2\sum_i m_i}\cdot\gamma^{-2P}=\gamma^2$,
using Eq.~\eqref{eq:P_count}.  The $u$ factors combine as
$u^P/u^{\sum_i m_i}=u^{-1}$,
again using Eq.~\eqref{eq:P_count}.  Defining $F(T/u)\equiv\prod_i f_i(T/u)$,
we obtain
\begin{equation}
    \Sigma_c^{(n_l)}
    \;\sim\; \frac{\gamma^2}{u}\,F\!\left(\frac{T}{u}\right)\,s^{2n_l},
    \label{eq:loop_scaling_final}
\end{equation}
confirming Eq.~\eqref{eq:loop_scaling}.  The $T/u$ dependence of individual
diagrams is encoded in $F(T/u)$.  For the non-local diagram of
Sec.~\ref{section:excluded}, one vertex contributes $f_i\sim u/T$, arising
from the static ($\omega=0$) flavor bubble whose Matsubara sum gives a factor
of $1/T$.  This yields an additional suppression of $s^2\cdot(u/T)\sim
\Theta_c/T$ relative to the one-loop result, consistent with
Eq.~\eqref{eq:excluded}.

As stated above, we have not accounted for combinatorial factors. These may contain factors of $N_f$. 
Since the vertices conserve the fermion flavor, each loop can contribute at most one factor of $N_f$ (or less, depending on the structure of the diagram).
This establishes $N_f s^2$ as the correct small parameter that justifies the loop expansion. 

\section{Full Four-Point Function}
\label{App:f4}
The two body vertex $\Gamma_2$ is derived from the connected four-point function 
\begin{align}
&\braket{f_{1}\left(\tau_{1}\right)f_{2}\left(\tau_{2}\right)\bar{f}_{2}\left(\tau'_{2}\right)\bar{f}_{1}\left(\tau'_{1}\right)}_{c,0} = \nonumber\\
&\braket{f_{1}\left(\tau_{1}\right)f_{2}\left(\tau_{2}\right)\bar{f}_{2}\left(\tau'_{2}\right)\bar{f}_{1}\left(\tau'_{1}\right)}_0 
    -\braket{f_{1}(\tau_{1})\bar{f}_{1}(\tau'_{1})}_0 \braket{f_{2}(\tau_{2})\bar{f}_{2}(\tau'_{2})}_0 
    +\braket{f_{1}(\tau_{1})\bar{f}_{2}(\tau'_{2})}_0 \braket{f_{2}(\tau_{2})\bar{f}_{1}(\tau'_{1})}_0. 
\end{align}
The two-point function is given by $\braket{f_{1}(\tau)\bar{f}_{2}(\tau')}_0=\delta_{\lambda_1,\lambda_2}\delta_{b_1,b_2}\tilde{G}_f^0(\tau-\tau')$, with $\tilde{G}_f^0(\tau-\tau')$ defined in the text. Below we give the expression for the full four-point function.
\subsection{Different Flavor-Orbitals}
The four point function with different flavor-orbital combinations $(\lambda_1,b_1)\ne(\lambda_2,b_2)$ is given by 
\begin{align}
&\braket{f_{1}\left(\tau_{1}\right)f_{2}\left(\tau_{2}\right)\bar{f}_{2}\left(\tau'_{2}\right)\bar{f}_{1}\left(\tau'_{1}\right)}_0 \nonumber\\
&=
\begin{cases}
A_1\exp\left[-E_{+}\left(\tau_{1}-\tau'_{1}+\tau_{2}-\tau'_{2}\right)-U\left(\min\{\tau_1,\tau_2\}-\max\{\tau'_1,\tau'_2\}\right)\right] & \left(\tau_{1},\tau_{2}\right)>\left(\tau'_{1},\tau'_{2}\right)\;\;\left(a\right)\\
A_2\exp\left[-E_{-}\left(\tau'_{1}-\tau_{1}+\tau'_{2}-\tau_{2}\right)-U\left(\min\{\tau'_1,\tau'_2\}-\max\{\tau_1,\tau_2\}\right)\right] & \left(\tau'_{1},\tau'_{2}\right)>\left(\tau_{1},\tau_{2}\right)\;\;\left(b\right)\\
-A_3 \tilde{G}_{ins}\left(\tau_{2}-\tau'_{1}\right)\tilde{G}_{ins}\left(\tau_{1}-\tau'_{2}\right) & \left(\tau_{1},\tau'_{2}\right)\gtrless\left(\tau'_{1},\tau_{2}\right)\;\;\left(c\right)\\
A_4\tilde{G}_{f}^{0}\left(\tau_{2}-\tau'_{2}\right)\tilde{G}_{f}^{0}\left(\tau_{1}-\tau'_{1}\right) & \left(\tau_{2},\tau'_{2}\right)\gtrless\left(\tau'_{1},\tau_{1}\right)\;\;\left(d\right)
\end{cases}
\end{align}
where $E_\pm$ are the single-body doublon like and holon like excitation energies as defined in the main text. $\tilde{G}_{ins}$ and $\tilde{G}_{f,0}$ are the flavor-flip instanton propagator and single-body $f$ propagator defined in the main text. $(a,b)>(c,d)$ denotes $\min\{a,b\}>\max\{c,d\}$. The combinatorial factors are given by $A_1=\frac{\left(1-n\right)\left(1-n-N_{f}^{-1}\right)}{1-N_{f}^{-1}}$, $A_2=\frac{n\left(n-N_{f}^{-1}\right)}{1-N_{f}^{-1}}$, $A_3=\frac{n\left(1-n\right)}{1-N_{f}^{-1}}$ and
$A_4=\frac{1}{1-N_f^{-1}}\begin{cases}
1-N_{f}^{-1}n^{-1} & \sum_{i}\text{sgn}\left(\tau_{i}-\tau'_{i}\right)=-2
\\
1 & \sum_{i}\text{sgn}\left(\tau_{i}-\tau'_{i}\right)=0\\
1-N_{f}^{-1}\left(1-n\right)^{-1} & \sum_{i}\text{sgn}\left(\tau_{i}-\tau'_{i}\right)=2\end{cases}$, with $n=\frac{1}{2}+\frac{\nu}{N_f}$ being the fraction of filled $f$ flavor-orbitals. In the limit of large $N_f$ these simplify to $A_1=\left(1-n\right)^2$, $A_2=n^2$, $A_3=n\left(1-n\right)$ and
$A_4=1$.

In Matsubara frequencies, expressing $\braket{f_{1}\left(\omega_{1}\right)f_{2}\left(\omega_{2}\right)\bar{f}_{2}\left(\omega'_{2}\right)\bar{f}_{1}\left(\omega'_{1}\right)}_0\equiv2\pi\delta(\omega_1+\omega_2-\omega'_1-\omega'_2)\mathcal{G}_{\uparrow\downarrow}(\omega_1,\omega_2;\omega_1',\omega_2')$, we find
\begin{align}
    &\mathcal{G}(\omega_1,\omega_2;\omega_1',\omega_2')= \nonumber\\
    &\frac{A_1}{2\left(U-\delta\mu\right)-i\left(\omega_{1}+\omega_{2}\right)}\left(\frac{1}{E_{+}-i\omega_{1}}+\frac{1}{E_{+}-i\omega_{2}}\right)\left(\frac{1}{E_{+}-i\omega'_{1}}+\frac{1}{E_{+}-i\omega'_{2}}\right) \nonumber\\
    +&\frac{A_2}{2\left(U+\delta\mu\right)+i\left(\omega'_{1}+\omega'_{2}\right)}\left(\frac{1}{E_{-}+i\omega_{1}}+\frac{1}{E_{-}+i\omega_{2}}\right)\left(\frac{1}{E_{-}+i\omega'_{1}}+\frac{1}{E_{-}+i\omega'_{2}}\right) \nonumber\\
    -&A_3\delta_{\omega_{1},\omega'_{2}}G_{ins}\left(\omega_{2}\right)G_{ins}\left(\omega_{1}\right)+\frac{A_3}{\left(E_{+}-i\omega_{1}\right)\left(E_{+}-i\omega'_{1}\right)\left(E_{+}-i\omega'_{2}\right)}+\frac{A_3}{\left(E_{-}+i\omega_{1}\right)\left(E_{-}+i\omega'_{1}\right)\left(E_{-}+i\omega'_{2}\right)} \nonumber\\
    +&\frac{A_3}{\left(E_{+}-i\omega_{1}\right)\left(E_{-}+i\omega'_{1}\right)}\left(\frac{1}{E_{+}-i\omega'_{2}}+\frac{1}{E_{-}+i\omega_{2}}\right)+\frac{A_3}{\left(E_{+}-i\omega_{2}\right)\left(E_{-}+i\omega'_{2}\right)}\left(\frac{1}{E_{+}-i\omega'_{1}}+\frac{1}{E_{-}+i\omega_{1}}\right) \nonumber\\
    +&\frac{1}{i\left(\omega_{1}-\omega'_{1}\right)}\left[\frac{A_1}{\left(E_{+}-i\omega_{1}\right)\left(E_{+}-i\omega'_{2}\right)}-\frac{A_1}{\left(E_{+}-i\omega'_{1}\right)\left(E_{+}-i\omega_{2}\right)} +\frac{A_2}{\left(E_{-}+i\omega'_{1}\right)\left(E_{-}+i\omega_{2}\right)} \right. \nonumber\\
    &\qquad\qquad\qquad -\frac{A_2}{\left(E_{-}+i\omega_{1}\right)\left(E_{-}+i\omega'_{2}\right)} +\frac{A_3}{\left(E_{+}-i\omega_{1}\right)\left(E_{-}+i\omega_{2}\right)}-\frac{A_3}{\left(E_{+}-i\omega'_{1}\right)\left(E_{-}+i\omega'_{2}\right)} \nonumber\\
    &\qquad\qquad\qquad \left. +\frac{A_3}{\left(E_{-}+i\omega'_{1}\right)\left(E_{+}-i\omega'_{2}\right)}-\frac{A_3}{\left(E_{-}+i\omega_{1}\right)\left(E_{+}-i\omega_{2}\right)} \right]
\end{align}

Physically, time orderings $(a),(b)$ correspond to a double excitation by injecting two electrons/holes to a single site. Time ordering $(c)$ correspond to a generalized flavor flip, where at time $\tau_1'\sim\tau_2$ an electron of flavor-orbital $(\lambda_1,b_1)$ is added and an electron of flavor-orbital $(\lambda_2,b_2)$ is taken out of the site, while the inverse occurs at $\tau_2'\sim\tau_1$. Lastly, process $(d)$ is equivalent to the disconnected correlator, of two single-body excitations occurring at non-overlapping times, with a modified combinatorial factor due to classical conditional probability. In the large $N_f$ limit $(c)$ is the only non-vanishing contribution to the connected correlator at long time differences.

In real frequencies, only the term proportional to $\delta_{\omega_1,\omega'_2}$ contributes to the quasiparticle lifetime at one-loop order, therefore we consider this term in the text (and its generalizations at higher multi-point functions). 

For the susceptibilities, calculated directly at time, the same term turns out to be the relevant one, giving a $\frac{1}{T}$ divergence of the flavor-susceptibility at one-loop order.

\subsection{Same Flavor-Orbitals}

For completeness we hereby give the expression for the four-point function at identical flavor-orbitals $(\lambda_1,b_1)=(\lambda_2,b_2)$. 
We have
\begin{align}
&\braket{f_{1}\left(\tau_{1}\right)f_{1}\left(\tau_{2}\right)\bar{f}_{1}\left(\tau'_{2}\right)\bar{f}_{1}\left(\tau'_{1}\right)}_0 \nonumber\\
&=
\begin{cases}
\frac{1+\sgn(\tau_{1}-\tau'_{1})\sgn(\tau_{2}-\tau'_{2})}{1+(1-2n)\sgn(\tau_{1}-\tau'_{1})}\tilde{G}_{f}^{0}\left(\tau_{2}-\tau'_{2}\right)\tilde{G}_{f}^{0}\left(\tau_{1}-\tau'_{1}\right) & \left(\tau_{2},\tau'_{2}\right)\gtrless\left(\tau_{1},\tau'_{1}\right)\\
-\frac{1+\sgn(\tau_{2}-\tau'_{1})\sgn(\tau_{1}-\tau'_{2})}{1+(1-2n)\sgn(\tau_{2}-\tau'_{1})}\tilde{G}_{f}^{0}\left(\tau_{1}-\tau'_{2}\right)\tilde{G}_{f}^{0}\left(\tau_{2}-\tau'_{1}\right) & \left(\tau_{1},\tau'_{2}\right)\gtrless\left(\tau_{2},\tau'_{1}\right)
\end{cases}.
\end{align}
Physically this amounts to two consecutive electron-like or two consecutive holon-like excitations. Any other process is forbidden by fermionic statistics.

In Matsubara frequencies, expressing $\braket{f_{1}\left(\omega_{1}\right)f_{1}\left(\omega_{2}\right)\bar{f}_{1}\left(\omega'_{2}\right)\bar{f}_{1}\left(\omega'_{1}\right)}_0\equiv2\pi\delta(\omega_1+\omega_2-\omega'_1-\omega'_2)\mathcal{G}_{\uparrow\uparrow}(\omega_1,\omega_2;\omega_1',\omega_2')$, we find
\begin{align}
    \mathcal{G}(\omega_1,\omega_2;\omega_1',\omega_2')=\frac{1}{i\left(\omega_{1}-\omega'_{1}\right)}&\left[ \frac{1-n}{\left(E_{+}-i\omega_{1}\right)\left(E_{+}-i\omega'_{2}\right)}-\frac{1-n}{\left(E_{+}-i\omega'_{1}\right)\left(E_{+}-i\omega_{2}\right)}  \right. \nonumber\\
    &\left. +\frac{n}{\left(E_{-}+i\omega'_{1}\right)\left(E_{-}+i\omega_{2}\right)} -\frac{n}{\left(E_{-}+i\omega_{1}\right)\left(E_{-}+i\omega'_{2}\right)} \right]\nonumber\\
    &- \left(\omega_1\leftrightarrow\omega_2\right) .
\end{align}

\section{Large $N$ Vertex Resummation}
\label{App:Large_N}
Consider the $m-1$ loops contribution to the self energy from inserting a single $\Gamma^{(m)}$ vertex. Denoting $x=(\tau,\R)$ we have
\begin{align}
    &\sum_{\R,\R'}\int_{\tau,\tau'}\tilde{G}_c^0(x'_0-x) \tilde{\Sigma}_{m-1}(x-x')\tilde{G}_c^0(x'-x_0) \nonumber\\
    &= \frac{1}{(m!)^2}\int_{\{\tau_i\},\{\tau'_i\}}\sum_{\substack{\{\lambda_i,a_i,\R_i\}\\\{\lambda'_i,a'_i,\R'_i\}}} \Gamma^{(m)}_{\substack{1\cdots m \\ 1'\cdots m'}}\left[\braket{\bar{c}_0c_{0'}\bar{c}_m \cdots\bar{c}_1 c_{1'}\cdots c_{m'} }-\braket{c_{0'}\bar{c}_0}\braket{\bar{c}_m \cdots\bar{c}_1 c_{1'}\cdots c_{m'}}\right]\nonumber\\
    &= \frac{1}{m!}\int_{\{\tau_i\},\{\tau'_i\}}\sum_{\{\lambda_i\}}\sum_{\substack{\{a_i,\R_i\}\\\{a'_i,\R'_i\}}} \Gamma^{(m)}_{\substack{1\cdots m \\ 1'\cdots m'}}\left[\braket{\bar{c}_0c_{0'}\bar{c}_m \cdots\bar{c}_1 c_{1'}\cdots c_{m'}}-\braket{c_{0'}\bar{c}_0}\braket{\bar{c}_m \cdots\bar{c}_1 c_{1'}\cdots c_{m'}}\right] \nonumber\\
    &\approx \frac{1}{m!}\int_{\{\tau_i\},\{\tau'_i\}}\sum_{\lambda_1\ne\lambda_2\cdots\ne\lambda_m}\sum_{\substack{\{a_i,\R_i\}\\\{a'_i,\R'_i\}}} \Gamma^{(m)}_{\substack{1\cdots m \\ 1'\cdots m'}}\left[\braket{\bar{c}_0 c_{0'}\bar{c}_m \cdots\bar{c}_1 c_{1'}\cdots c_{m'} }-\braket{c_{0'}\bar{c}_0}\braket{\bar{c}_m \cdots\bar{c}_1 c_{1'}\cdots c_{m'}}\right] \nonumber\\
    &=- \frac{1}{(m-1)!}\int_{\{\tau_i\},\{\tau'_i\}}\sum_{\lambda_1\ne\lambda_2\cdots\ne\lambda_m}\delta_{\lambda_0,\lambda_m}\sum_{\substack{\{a_i,\R_i\}\\\{a'_i,\R'_i\}}} \Gamma^{(m)}_{\substack{1\cdots m \\ 1'\cdots m'}} \braket{c_{0'}\bar{c}_m}\braket{c_{m'}\bar{c}_0}\prod_{i=1}^{m-1}
    \braket{\bar{c}_{i}c_{i'}} .
\end{align}
 For the transition from the second to the third row we used the fact that $\Gamma^{(m)}$ is non-vanishing only if $\{\lambda_i\}$ and $\{\lambda_i'\}$ are equal up to a permutation of indices allowing us to assume $\lambda_i'=\lambda_i$, multiplying by a factor of $m!$ the size of the permutation group. In the third transition we took the large $N_\lambda$ limit to neglect configurations with $\lambda_i=\lambda_j$ for $i\ne j$. For the fourth transition we used the fact that the expression in the square bracket vanishes if none of the summed flavors $\lambda_i$ is equal $\lambda_0$ to assume specifically that $\lambda_m=\lambda_0$. Multiplying by a factor of $m$ accounts for the multiplicity of choices for any $\lambda_i=\lambda_0$. These steps  allow us to identify
\begin{align}
    &\tilde{\Sigma}_{m-1}(x_m-x'_m) \nonumber\\
    &= -\frac{1}{(m-1)!}\int \prod_{i=1}^{m-1}\left(d\tau_i d\tau_i'\right) \sum_{\lambda_1\ne\lambda_2\cdots\ne\lambda_{m-1}}\sum_{\substack{\{a_i,\R_i\}\\\{a'_i,\R'_i\}}} \Gamma^{(m)}_{\substack{1\cdots m \\ 1'\cdots m'}} \prod_{i=1}^{m-1}
    \braket{\bar{c}_{i}c_{i'}} \nonumber\\
    &= -\sum_{\R''}\hat{\gamma}^\dagger(\R_m-\R'')\left[ \frac{\sum_{\lambda_1\ne\lambda_2\cdots\ne\lambda_{m-1}}}{(m-1)!}\int \prod_{i=1}^{m-1}\left(d\tau_i d\tau_i'\right)   \sum_{\{b_i\}\{b'_i\}} \braket{f_1\cdots f_m \bar{f}_{m'} \cdots \bar{f}_{1'}}_{c,0} \sum_{\substack{\{a_i,\R_i\}\\\{a'_i,\R'_i\}}}\prod_{i=1}^{m-1}
    \braket{(\hat{\gamma}^\dagger_i\bar{c}_{i})(\hat{\gamma}_{i'}c_{i'})}
    \right]\hat{\gamma}(\R''-\R'_m) \nonumber\\
    &\approx  -\sum_{\R''}\hat{\gamma}^\dagger(\R_m-\R'')\left[ \frac{1}{(m-1)!}\int \prod_{i=1}^{m-1}\left(d\tau_i d\tau_i'\right)  \braket{f_1\cdots f_m \bar{f}_{m'} \cdots \bar{f}_{1'}}_{c,0} \prod_{i=1}^{m-1} \underset{=\tilde{\mathcal{I}}_0(\tau_i'-\tau_i)}{\underbrace{\sum_{b_i,\lambda_i} \sum_{\substack{a_i,\R_i\\a'_i,\R'_i}}
    \braket{(\hat{\gamma}^\dagger_i\bar{c}_{i})(\hat{\gamma}_{i'}c_{i'})}}}
    \right]\hat{\gamma}(\R''-\R'_m),
\end{align}
where we used the fact that given that all $\lambda_i$ are distinct the $f$ correlator is non-vanishing only if $\forall i:b_i=b'_i$, as the $f$ correlators are diagonal in orbital space. Given that this condition is satisfied, the $f$ correlator is independent of the values of $\lambda_i$ and $b_i$ due to the enhanced $SU(N_f)$ symmetry of the $f$ action (with $N_f =N_\lambda \times N_b$).
As a last step we use the relations between $\Sigma$ and $D$ to express
\begin{equation}
    \tilde{D}_{m-1}(\tau_m-\tau'_m)\approx -\frac{1}{(m-1)!}\int \prod_{i=1}^{m-1}\left(d\tau_i d\tau_i'\right)  \braket{f_1\cdots f_m \bar{f}_{m'} \cdots \bar{f}_{1'}}_{c,0} \prod_{i=1}^{m-1}\tilde{\mathcal{I}}_0(\tau'_i-\tau_i),
\end{equation}
where $f_i\equiv f_{\R,\lambda_i,b_i}(\tau_i)$ and $\bar{f}_{i'}\equiv \bar{f}_{\R,\lambda_i,b_i}(\tau'_i)$ are in the same flavor-orbital.

\subsection{analytical continuation of convolution}
\label{App:analytical_conv}
Consider a function $g(\omega)$ and its spectral function $\mathcal{A}_g(\omega)=-\frac{1}{\pi}\operatorname{Im}\left[g(\omega+i0_+)\right]$ admitting Lehmann's representation, such that we can express $g(z)=\int d\omega' \frac{\mathcal{A}_g(\omega')}{z-\omega'}$ (with $z$ being an abitrary complex number). We find 
\begin{equation}
    \frac{2}{i\Omega}\circledast g(i\Omega) = \int \frac{d\Omega'}{2\pi}\frac{2}{i(\Omega-\Omega')}\int d\omega'' \frac{\mathcal{A}_g(\omega'')}{i\Omega'-\omega''}
    =\int d\omega'' \frac{\mathcal{A}_g(\omega'')}{i\Omega-\omega''}\sgn(\omega'').
\end{equation}

After Wick rotating ($i\Omega\rightarrow\omega+i0^+$) we have
\begin{equation}
\label{eq:F11}
    \int d\omega'' \frac{A_g(\omega'')}{\omega-\omega''+i0^+}\sgn(\omega'')~.
\end{equation}
Taking the imaginary part we find
\begin{equation}
    \frac{2}{i\Omega}\circledast g(i\Omega) 
    \xrightarrow[\text{Im. part}]{i\Omega\rightarrow\omega+i\eta}
    \sgn(\omega)\operatorname{Im}\left[ g(\omega+i\eta)\right],
\end{equation}
whereas the real-part is given by the convolution
\begin{equation}
    \mathcal{P}\left[\int d\omega'' \frac{A_g(\omega'')}{\omega-\omega''}\sgn(\omega'')\right]=\mathcal{P}\left(\frac{2\pi}{\omega}\right)\circledast\left[A_g(\omega)\sgn(\omega)\right]
\end{equation}
as can be seen from eq.\eqref{eq:F11} or using Kramers-Kronig relations.

\section{Time Ordering of Single Site Correlators}
\label{App:time_ordering}

Consider the correlator
\begin{equation}
    \braket{f_1\cdots f_m \bar{f}_{m'} \cdots \bar{f}_{1'}}_{0}\equiv \frac{1}{N_{\text{states}}}\sum_{i=1}^{N_{\text{states}}}\braket{\Omega _i |\mathcal{T}\left\{\hat{f}_1(\tau_1)\cdots \hat{f}_m(\tau_m) \hat{f}^\dagger_m(\tau'_m) \cdots \hat{f}^\dagger_1(\tau'_1)\right\}|\Omega _i},
    \label{eq:D_1}
\end{equation}
where $\ket{\Omega_i}$ are states in the ground-state manifold. In our case this is simply a set of (orthogonal) states with the ground-state charge $\delta n\ket{\Omega_i}=\nu\ket{\Omega_i}$. Notice that all the inserted creation and annihilation operators create energy eigenstates (as all number eigenstates are energy eigenstates). We can therefore separate between the dynamical (time-dependent) part of Eq.~\eqref{eq:D_1} and the time independent numerical prefactors (that include the sign) from time ordering and matrix elements with the different ground states to write 
\begin{equation}
    \braket{f_1\cdots f_m \bar{f}_{m'} \cdots \bar{f}_{1'}}_{0}=\exp\left[-F(\tau_1\cdots\tau_m,\tau'_1\cdots\tau'_m)\right] \frac{\sum_{i=1}^{N_{\text{states}}}}{N_{\text{states}}}\lim_{\epsilon\rightarrow0}\braket{\Omega _i |\mathcal{T}\left\{\hat{f}_1(\epsilon\tau_1)\cdots \hat{f}_m(\epsilon\tau_m) \hat{f}^\dagger_m(\epsilon\tau'_m) \cdots \hat{f}^\dagger_1(\epsilon\tau'_1)\right\}|\Omega _i}.
\end{equation}
Here $F$ is a real-valued function encoding the time-dependence of the correlator.
Notice that now the correlator on the RHS is static, where the time arguments of the operators only determine the ordering.
We can choose $\ket{\Omega_i}$ to all be Slater determinants, and therefore use Wick's theorem to compute the static correlator. 
The sign is given by $\prod_i\operatorname{sgn}(\tau_i-\tau'_i)$ (in the main text we took a minus sign out, therefore the sign functions are flipped).

\section{Excluded Diagram}
\label{App:excluded}
The contribution in real space to $\left[D_\lambda(\R-\R',\omega)\right]_{b,b'}$ is
\begin{multline}
    -\left[1-\left(\frac{2\nu}{N_f}\right)^2\right]^2\frac{\sum_{\R,\R'}}{N_{U.C.}^2}\frac{\sum_{\omega_1\cdots\omega_3}}{\beta^{3}}\sum_{\alpha_1\cdots\alpha_3}\beta\delta_{\omega,\omega_1}\left( \beta\delta_{\omega_2,-\omega_3} \right)^2\delta_{\alpha_1,\alpha_2}\delta_{\alpha,\alpha_3} 
    \left( \frac{u}{u^2-(i\omega+\delta\mu)^2} \frac{u}{u^2-(i\omega_3+\delta\mu)^2} \right)^2\\ \times  
    \prod_{j=1}^3 
    \sum_{\R_j,\R'_j}\left[\hat{\gamma}\left(\R-\R_j\right)G_{c,\lambda_j}\left( \R_j-\R'_j,i\omega_j\right)\hat{\gamma}^\dagger\left(\R'_j-\R'\right)\right]_{b_j,b'_j}~,
 \end{multline}
with $\beta=T^{-1}$ being the inverse temperature and $\alpha_j=(\lambda_j,b_j,b_j')$ a super index of flavor and $f$-orbital indices for incoming and outgoing states. Orbital indices of $c$-electrons are suppressed and are all summed over as dictated by matrix multiplication rules for $\hat{\gamma} G_c\hat{\gamma}^\dagger$. approximating Matsubara summation by integral in the low temperature limit
\begin{align}
    \overset{\beta\rightarrow\infty}{\Rightarrow}-\beta&\left[1-\left(\frac{2\nu}{N_f}\right)^2\right]^2\left(\frac{u}{u^2-(i\omega+\delta\mu)^2}\right)^2 \frac{\sum_{\R,\R'}}{N_{U.C.}^2} \sum_{\alpha_1}\sum_{\R_1,\R'_1}
    \left[\hat{\gamma}\left(\R-\R_1\right)G_{c,\lambda_1}\left( \R_1-\R'_1,i\omega\right)\hat{\gamma}^\dagger\left(\R'_1-\R'\right)\right]_{b_1,b'_1}\nonumber\\ 
    &
    \begin{aligned}
    \times  \int_{\omega'}\left(\frac{u}{u^2-(i\omega'+\delta\mu)^2}\right)^2 &\sum_{\R_2,\R'_2}
    \left[\hat{\gamma}\left(\R-\R_2\right)G_{c,\lambda_1}\left( \R_2-\R'_2,-i\omega'\right)\hat{\gamma}^\dagger\left(\R'_2-\R'\right)\right]_{b_1,b'_1}\\
    \times&\sum_{\R_3,\R'_3}
    \left[\hat{\gamma}\left(\R-\R_3\right)G_{c,\lambda}\left( \R_3-\R'_3,i\omega'\right)\hat{\gamma}^\dagger\left(\R'_3-\R'\right)\right]_{b,b'},
    \end{aligned}
\end{align}

By doing a Fourier transform of the spatial coordinates, we find the contribution to $\left[D_\lambda(\k,i\omega)\right]_{b,b'}$ given by Eq.~\eqref{eq:excluded_full},\eqref{eq:Theta_c} in the main text.

\section{Flavor Susceptibility}
\label{App:susceptibility}

As we showed in the main text, the Curie temperature at charge neutrality at the flat-chiral limit is maximal at $\q=0$ and is equal for any type of flavor order. To examine the effects of bandwidth ($M\ne0$) and deviation from the chiral limit ($v_\star'\ne0$) we plot the eigenvalues of the matrix $\hat{\Theta}^{(\lambda,\lambda')}(\q)$, corresponding to the Curie temperatures of the different orders, in Fig.~\ref{fig:T_curie}. We find that finite $M$ and finite $v_\star'$ independently lift part of the degeneracies, but the degeneracy between pairs of flavor diagonal and flavor off-diagonal order is only lifted in the presence of both finite $v_\star'$ and finite $M$. In this case the leading instability is in a flavor off-diagonal susceptibility (i.e. towards an intervalley coherent order) in agreement with the predicted zero-temperature order~\cite{bultinck2020ground}.

In addition to splitting degeneracies, we find that a finite $M$ leads to a logarithmic divergence of $\hat{\Theta}^{(\lambda,\lambda')}(q)$. 
The logarithmic divergence originates from the quadratic band touching of the c-electron dispersion at zero energy, which arises when $M\ne0$. 
In reality this divergence is cut off either by finite temperature or by the finite lifetime of the low energy excitations (obtained when the internal propagators are calculated beyond the Hubbard-I approximation). 
In Fig. \ref{fig:T_curie} we used $T=s^2 u/2$ to regularize the divergence. 


\begin{figure}[H]
    \centering
    \includegraphics[width=0.7\linewidth]{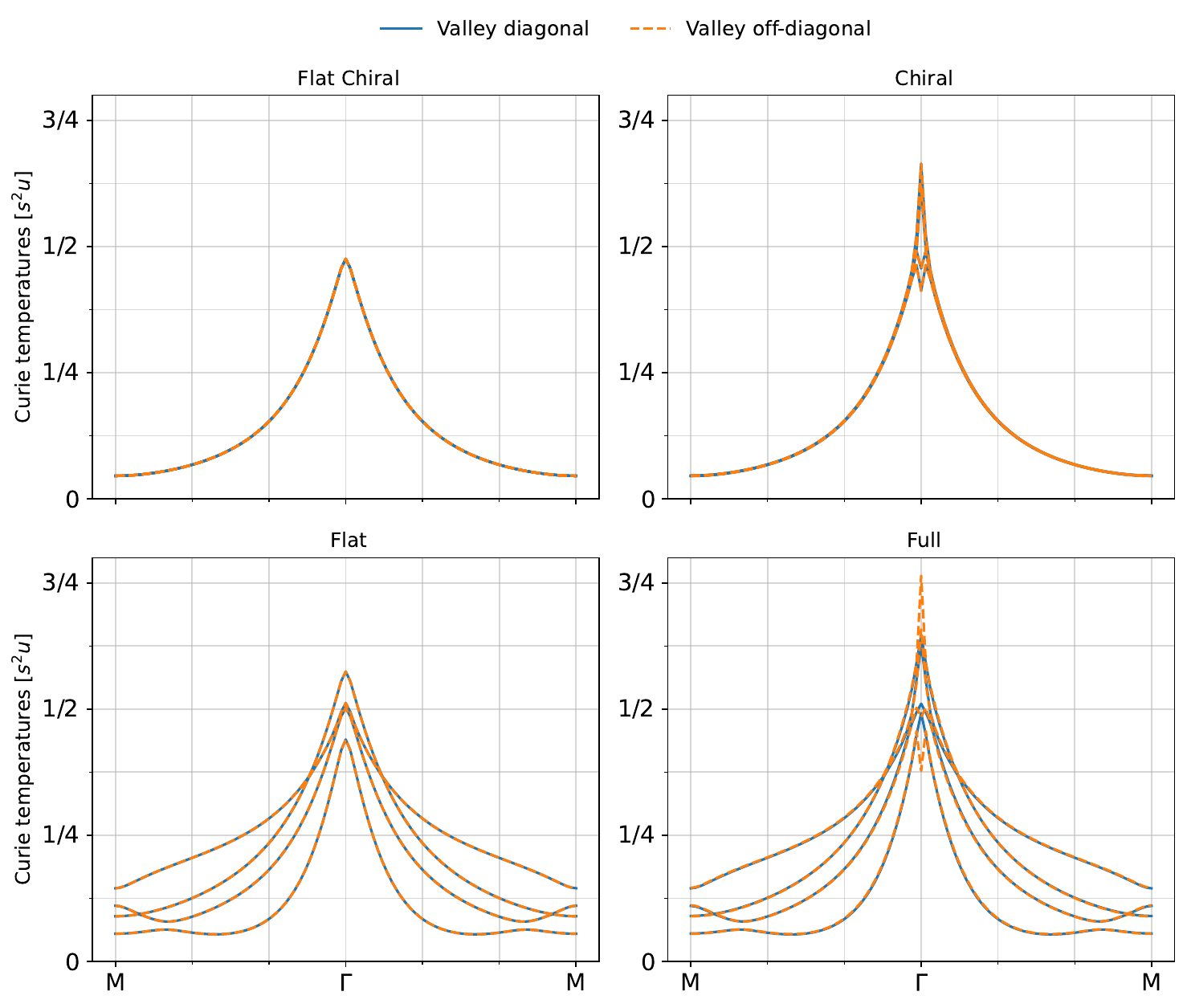}
    \caption{The eigenvalues of the Curie temperature matrix $\hat{\Theta}^{(\lambda,\lambda')}(q)$ as function of momentum, in units of the ordering scale $s^2u$. We plot the results separately for valley diagonal ($\tau=\tau'$) and valley off-diagonal ($\tau\ne\tau'$) susceptibilities, where $\lambda=(s,\tau)$ is the flavor index. We show the results for four different scenarios: (a) flat-chiral ($M=v_\star'=0$), (b) chiral ($M\ne0$, $v_\star'=0$), (c) flat ($M=0$, $v_\star'\ne0$), (d) full model ($M\ne0$, $v_\star'\ne0$). $\hat{\Theta}^{(\lambda,\lambda')}(q)$ is calculated according to Eq.~\eqref{eq:Theta_c}, by doing a discrete Matsubara summation corresponding to a temperature of $T=\frac{s^2u}{2}$.}
    \label{fig:T_curie}
\end{figure}
\end{widetext}
\bibliography{ref}
\end{document}